\documentstyle[a4wide,amssymb,graphicx,aps]{revtex}

\begin{document}

\title{Random tiling quasicrystals in three dimensions}

\author{W. Ebinger, J. Roth, H.-R. Trebin,\\
Institut f\"ur Theoretische und Angewandte Physik\\
Universit\"at Stuttgart\\
Pfaffenwaldring 57\\
70550 Stuttgart, Germany}

\date{\today}

\maketitle

\begin{abstract}
  Three-dimensional icosahedral random tilings with rhombohedral cells
  are studied in the semi-entropic model. We introduce a global energy
  measure defined by the variance of the quasilattice points in the
  orthogonal space. The internal energy, the specific heat, the
  configuration entropy and the sheet magnetization (as defined by
  Dotera and Steinhardt [Phys.~Rev.~Lett.\ {\bf 72} (1994) 1670]) have
  been calculated. The specific heat shows a \textsc{Schottky} anomaly
  which might indicate a phase transition from an ordered quasicrystal
  to a random tiling. But the divergence with the sample size as well as
  the divergence of the susceptibility are too small to pinpoint the
  phase transition conclusively. The self-diffusion coefficients
  closely follow an \textsc{Arrhenius} law, but show plateaus at
  intermediate temperature ranges which are explained by energy
  barriers between different tiling configurations due to the harmonic
  energy measure. There exists a correlation between the temperature
  behavior of the self-diffusion coefficient and the frequency of
  vertices which are able to flip (simpletons).  Furthermore we
  demonstrate that the radial distribution function and the radial
  structure factor only depend slightly on the 
  random tiling configuration. Hence, radially symmetric pair
  potentials lead to an energetical equidistribution of all
  configurations of a canonical random tiling ensemble and do not
  enforce matching rules.

\end{abstract}

\section{Introduction}
The stability of quasicrystals has been a riddle since their discovery
in 1982. Do they form stable or metastable states? Are they stable
only at high temperatures? Is the stability due to energetic or due to
entropic reasons? The random tiling model of quasicrystals is an
abstraction which deals with rigid tiles, thereby neglecting thermal
fluctuations and the phonon degrees of freedom. The only dynamic
process is the local rearrangement of tiles, called ``flips'',
``umklapps'', ``zipper'' moves, depending on the type of tiles and
symmetries. The random tiling model has been proposed by
Elser~\cite{elser85} and has been studied intensively in recent years.
The first author dealing with random rhombohedral tilings in three
dimensions was Tang~\cite{tang90} who was interested in diffuse
scattering and phason elastic constants. Strandburg~\cite{strandburg91}
calculated the configurational entropy. The state of the art of random
tilings was reviewed by Henley~\cite{henley91}.
Ebinger~\cite{ebinger91} studied random tilings at infinite
temperatures. 
Dotera and Steinhardt~\cite{dotera94} introduced the concept of sheet
magnetization as an order parameter to describe the randomness.
Interest in random tilings was renewed by Kalugin and
Katz~\cite{kalugin} through the new process of flip diffusion.  This 
property has been studied by Jari\'c and
S{\o}rensen~\cite{sorensen94,jaric95} at infinite temperature, and by
Joseph~\cite{joseph95} and G\"ahler~\cite{gaehler95} at finite
temperatures. Meanwhile for the energetic interaction of the tiles
many different sets of matching rules exist.

In the present work we first deal with quantities that allow to
characterize a random tiling. We are especially
interested in the atomic plane and line structures important for
example for ion channeling (Sec.~\ref{character}).  Energy measures
(Sec.~\ref{energy}) are discussed and a new method for generating
non-equilibrium quasicrystals (Sec.~\ref{construct}) is established.
The possibility of a random tiling equilibrium phase transition is
studied in Sec.~\ref{random}. Structure functions are introduced in
Sec.~\ref{structure}, and results concerning a special ``harmonic
energy measure'' as defined in Sec.~\ref{energy} are presented in
Sec.~\ref{results}. Conclusions are drawn in Sec.~\ref{conclude}.

\section{Definitions}\label{defs}

Quasicrystals can be described as cuts through higher-dimensional
periodic crystals. The additional dimensions are addressed as
orthogonal space $\mathbb{E}^{\perp}$. Quasilattice points
${\mathbf x}$ in the physical space $\mathbb{E}^{\parallel}$ can uniquely be
lifted to the higher-dimensional space
${\mathbf X}=\Pi_{{\parallel}}^{-1}{\mathbf x}$ and then be projected onto the
orthogonal space by ${\mathbf y} = \Pi_{\perp}({\mathbf X})$
($\Pi_{\parallel}+\Pi_{\perp} = id$ in a proper
normalization). The infinitely extended quasilattice thus is contracted
into a finite volume called ``acceptance domain'' or ``atomic
hypersurface''. The whole procedure of lifting and projecting into the
orthogonal space is called ``dualization'' ${\mathbf y} = \Pi_{\perp}
\circ \Pi_{\parallel}^{-1} {\mathbf x}$ of the quasilattice. The
higher-dimensional embedding generates new degrees of freedom in
addition to the ordinary phonons in periodic crystals, denoted
``phasons''.

The icosahedral quasilattice lifted into higher-dimensional space
forms a three-di\-men\-sional hypersurface, called
``de-Bruijn-hypersurface'' or ``Weiringia roof''~\cite{debruijn}. The
hypersurface fluctuates around an average hyperplane ${\mathbf
  h}({\mathbf x})={\mathbf h}_{0}+\varepsilon_{global}{\mathbf x}$.
The constant quantity
\begin{displaymath} 
\varepsilon_{global} :=\nabla_{\parallel} \otimes
{\mathbf h}({\mathbf x})=const
\end{displaymath}
is called ``global phason strain'' and describes the deviation of the
slope of the average hyperplane $\mathbf{h}(\mathbf{x})$ from the
slope of the physical space.  In the case of exact icosahedral
symmetry the average hyperplane is running parallel to the
physical space and thus $\varepsilon_{global}=0$.\\
Long-wavelength deviations from the hypersurface are denoted as ``phason
fluctuations'' and are described in a continuum picture by a phason
strain tensor
\begin{displaymath} 
\varepsilon(\mathbf{x}) := \nabla_{\parallel} \otimes
\mathbf{h}(\mathbf{x})\quad. 
\end{displaymath}
According to Henley~\cite{henley88} the fluctuations 
are  governed by a free energy depending quadratically on the
phason strain:
\begin{displaymath}
F=\int d^{d}\mathbf{x} \, \mathrm{trace}
\left(\varepsilon(\mathbf{x})\otimes\varepsilon(\mathbf{x})^{T}\right)\quad. 
\end{displaymath}
The dimension $d$ of physical space is equal to three for
icosahedral quasicrystals.

The present work deals with the three-di\-men\-si\-on\-al
Ammann-Kramer-Penrose tiling and its randomizations.  This tiling
consists of two different elementary cells, the ``oblate'' and
``prolate'' rhombohedron. Both of them appear in ten different
orientations in the tiling. The six rational linearly independent unit
edge vectors ${\mathbf t}_{i}$ may be defined as ${\mathbf
  t}_{\alpha} = \frac{\sqrt{\displaystyle 2}}{\displaystyle
  5}(\cos\frac{2}{5}\pi\alpha,\sin\frac{2}{5}\pi\alpha,\frac{1}{2})$ 
($\alpha$=0,$\dots$,4) and ${\mathbf t}_{5}$=(0,0,1).  The atomic
hypersurface is a rhombic triacontahedron of icosahedral symmetry. The tiling
exhibits, among others, a vertex representing lattice points where two
prolate and two oblate rhombohedra meet. It is denoted ``simpleton
vertex''. The outer shape of the cells, which touch this vertex, is a
rhombic dodecahedron. If the dodecahedron is kept fixed, two
possibilities exist to fill it with tiles. The exchange of one
configuration by the other is a ``flip''. The lattice point jumps a distance
which is 0.650 of the edge length of the rhombohedra
(Fig.~\ref{flip}). Any non-ordered space 
filling arrangement of rhombohedra without gaps or overlaps is called
a ``random tiling''.

\section{Random tiling characterization} 
\label{character}

\subsection{Variance} 
\label{variance}

Any random tiling may be characterized by the mean square deviation
of the point distribution from the center of mass in the orthogonal
space. 
The variance is defined by
\begin{equation}\label{vareq}
\Omega=\frac{1}{N}\sum^N_{j=1}\left|{\bf y}_j -
\frac{1}{N}\sum^N_{i=1}{\bf y}_i\right|^2 = \frac{1}{N}\sum^N_{j=1}|{\bf
y}_j - {\bf y}_0|^2\quad.
\end{equation}
$N$ is the number of quasilattice sites and ${\bf y}_0$ is the average
of the position vectors ${\bf y}_i$ of all dual quasilattice sites.
The bigger the phason fluctuations are the larger is the disorder and
therefore the random tiling parameter $\Omega$. The averaged value in
the pure entropic random tiling ensemble amounts to
1.73$\pm$0.01~\cite{tang90}. 

\subsection{Vertex frequency} 

The lattice points of a quasicrystal can be characterized by their
local environment. The number of rhombohedra adjacent to a lattice point in
an icosahedral quasilattice varies from 4 to 20, the number of edges
from 4 to 12. There are 24 ``canonical'' or ``allowed'' vertices, but
5450 vertices may occur in a random tiling. Vertices where two
identical rhombohedra meet are called ``crystallographic''. Without
them the number of vertices is reduced to 360~\cite{benabraham93}.

In the two-dimensional Penrose tiling each flip generates
non-canonical vertices. There the knowledge of the vertices suffices to
distinguish between ideal and random tilings. This statement is not
valid in three dimensions. There are some flips possible in the
rhombohedra tiling which only change the frequency of vertices
without introducing forbidden ones. If the degree of randomization has
reached a certain level the number of forbidden vertices is going to
rise rapidly.

\subsection{Sheet magnetization}\label{sheetmag}  

The frequency of the simpleton is 23.61\% in the ideal rhombohedra
tiling. The simpletons are arranged in two-dimensional layers
perpendicular to the two-fold symmetry axis. Due to the two
possibilities to pack the rhombohedra into the simpleton there are two
positions for the internal lattice point. The positions may be called
``up'' and ``down'' and attributed a spin value $S=\pm1$. In the ideal
rhombohedra tiling all the spins in a certain layer carry spin +1 or
--1. Dotera and Steinhardt~\cite{dotera94} have defined a sheet
magnetization of value 1 using a proper summation rule.
In a random tiling the sheets also exist, but the magnetization is
reduced since the spins in one layer are not all aligned.
%
%
The collection of all sheets can therefore be regarded as a
two-dimensional Ising model with a variable number of spins. There is,
however, no interaction leading to an 
Ising Hamiltonian. The sheet magnetization simply is being used as an
order parameter which may indicate an order-disorder transition. The
number of spins is not constant as in a standard Ising system. 
The susceptibility is given by 
\begin{displaymath}
\chi=\frac{1}{T}<N_D>\left( \left< \frac {M^2}{N^2_D}\right> -
\left< \frac{M}{N_D}\right> ^2\right)\quad.
\end{displaymath}
$N_D$ is the number of the dodecahedra and therefore the number of the
``spins'' in the patch. This equation replaces
the more cumbersome definition used in~\cite{dotera94}.

\subsection{Tube and slab structures}\label{plane} 

In the three-dimensional icosahedral rhombohedra tiling special tube
structures, denoted ``worms'', exist along two-fold
axes (Fig.~\ref{worm}). They are 
formed of stacks of rhombohedra which have one type of face in common,
perpendicular to the two-fold axis. In addition to the tube
structures there are slab structures perpendicular to a five-fold
direction. The slabs are formed by rhombohedra which have one type of
edge in common (parallel to the five-fold axis). The projection of the
slab structures onto the plane perpendicular to the appropriate
five-fold direction yields generalized Penrose tilings in two
dimensions. These tube and slab structures exist in ideal tilings as
well as in random tilings.


\subsection{Plane and chain density diagrams}\label{channeling} 

Atoms in crystals are arranged in lattice planes characterized by a
normal vector of rational components (the Miller indices), and in
lattice chains with a rational direction vector. This fact also holds
for quasicrystals. The lattice planes or chains of a fixed direction
in a crystal all are equivalent to one another in the simple cases (or
they fall into a small number of classes). This is not true for a
quasicrystal: For example, the density of atoms in a certain plane does
not only depend on the orientation of the plane but also on its
position. The basic theory can be found in Ref.~\cite{katz86}. The
plane and chain structures have been studied in
Ref.~\cite{kupke92,kupke93}. 

As described in Sec.\ \ref{defs} it is possible to lift the
quasilattice points into the higher-dimensional space and to project them
onto the orthogonal space. Points in the same physical plane are situated
in the same plane in the orthogonal space~\cite{katz86}, and
points on one chain in the physical space
will lie on one chain in the orthogonal space, in the
ideal tiling within the acceptance domain,
respectively.

To describe densities of atoms in the planes of our (primitively
decorated) quasicrystal, we use the function $A_s(\xi^{\perp})$. The
index $s$ shall indicate a two-, three- or five-fold symmetry
direction. $\xi^{\perp}$ is the coordinate of position of a plane in
the direction parallel to its normal and 
is the dual of the height $\xi^{\parallel}$
of the selected plane in the physical space (with respect to the
origin for example).  $A_{s}(\xi^{{\perp}})$ is proportional to the
area of the intersection plane at $\xi^{\perp}$ through the
triacontahedral acceptance domain in the ideal tiling, since the
density of points is 1 in the domain and 0 outside.  

For random tilings we have to use another definition. The area of
$A_s(\xi^{\perp})$ at the position $\xi^{\perp}$ in
$\mathbb{E}^{\perp}$ is proportional to the plane decoration
density $\rho_{F,s}(\xi^{\parallel}[\xi^{\perp}]) =:
\rho_{F,s}(\xi^{\parallel})$ of the corresponding plane
$\xi^{\parallel}=const$ in physical space. The index $F$ denotes
plane densities.

\begin{figure}
\caption{Plane structures in the ideal tiling. The left column contains
  the functions $A_{s}(\xi^{\perp})$, the right column displays the
  histograms of plane densities $H(\rho_{F,s})$.\label{planid}}
\vspace*{1cm}
\begin{center}
\includegraphics[width=12cm,angle=0]{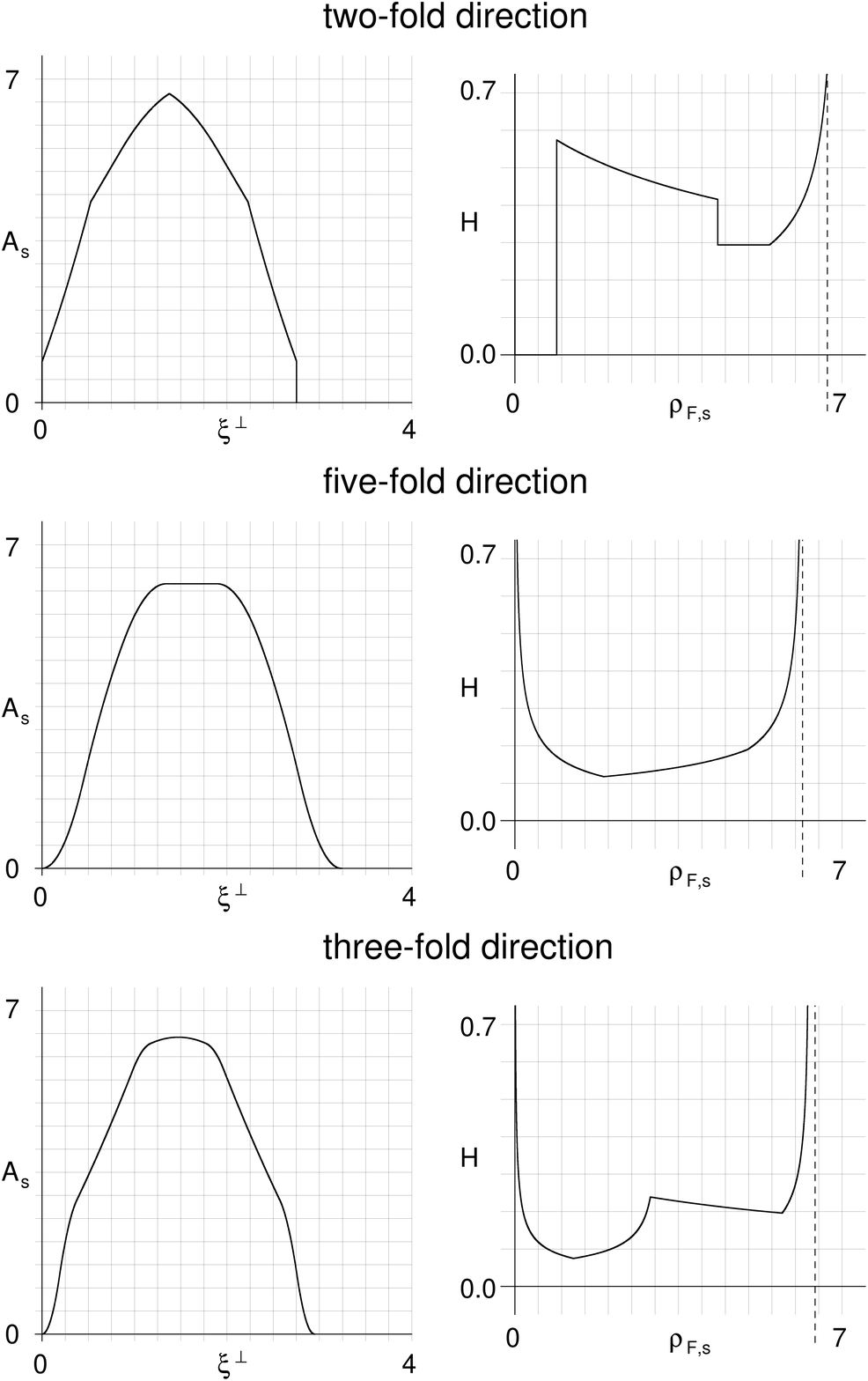}
\end{center}
\end{figure}

In the case of an ideal quasicrystal, the values of $A_s(\xi^{\perp})$
are between 0 and a maximal value for five-fold and three-fold
directions (Fig.~\ref{planid}). In the five-fold case the
function has a constant plateau since the corners of the intersection
polygon are formed by edges of the triacontahedron perpendicular to the
plane. In the two-fold case the minimal value of $A_s(\xi^{\perp})$ is
greater than zero, since the smallest intersection polygons coincide
with the surface-rhombi of the triacontahedron parallel to the 
intersection plane. To obtain the average density of atoms in a plane
one has to calculate the mean value of the area
of the intersection polygons. It is equivalent to the volume of the
triacontahedron divided by the diameter $\Delta_s$ of the
triacontahedron along the direction $s$. Then the mean value of the
plane density is: 
\begin{displaymath}
{\bar \rho}_{F,s} \propto {\bar A}_s = \frac{V_{Tria}}{\Delta_s}\quad.
\end{displaymath}

For channeling experiments one is interested in the frequency
$H(\rho_{F,s})$ of planes perpendicular to $s$ with a plane density
$\rho_{F,s}$ in the interval ${\delta}\rho_{F,s}$. Wherever the
intersection function $A_s(\xi^{\perp}) \propto
\rho_{F,s}(\xi_{\parallel})$ is flat, there are many
$\xi^{\parallel}[\xi^{\perp}]$ values for $\rho_{F,s}$ inside of
${\delta}\rho_{F,s}$. If the slope is large, the number of planes with
this density is small.
The densities $H(\rho_{F,s})$ are obtained
from the values $\rho_{F,s}$ by a subdivision of the $\rho_{F,s}$ axis into
intervals $\delta \rho_{F,s}$ of equal length and calculating for each
interval ${\delta}\rho_{F,s}$ the length of the inverse image on the
$\xi^{\perp}$-axis.
Since the
triacontahedron is a convex polyhedron and centrosymmetric, the
function $A_s(\xi^{\perp})$ is monotonously growing in one half
of the interval $\Delta_{s}$ and monotonously decreasing in the
other. Therefore 
one can calculate the first derivative of the inverse image (that means
${\delta}\rho_{F,s}\to 0$) in one half and set it proportional to
$H(\rho_{F,s})$ which hence is $\xi^{\perp\prime}(A_{s})$. Since
$A_s(\xi^{\perp})$ has horizontal lines and kinks, $H(\rho_{F,s})$
possesses poles and jumps. 

Similarly as the plane structures one also finds chains in the
quasicrystal. The direction again is denoted by $s$, the position in the
perpendicular space now is characterized by a two-dimensional parameter
$\{\eta^{\perp},\zeta^{\perp}\}=const$ with its counterpart in the
physical space at $\{\eta^{\parallel},\zeta^{\parallel}\}=const$. The
chain density $\rho_{l,s}(\eta^{\perp},\zeta^{\perp})$ at
$[\eta^{\perp},\zeta^{\perp}]$ is proportional to the intersection
distance $w_s(\eta^{\perp},\zeta^{\perp})$ of the dual line through
the triacontahedron. The index $l$ denotes line densities.

To obtain the average chain density for a given direction $s$ one has
to calculate the average intersection length ${\bar w}_s$. It is
equivalent to the volume of the triacontahedron divided by the area
$\Phi_{Tria,s}$ of the triacontahedron projected onto the
$\eta^{\perp}\zeta^{\perp}$-plane:
\begin{displaymath}
{\bar \rho}_{l,s} \propto {\bar w}_s = \frac{V_{Tria}}{\Phi_{Tria,s}}\quad.
\end{displaymath}

It is also possible to calculate frequencies of the chain densities.
The $\rho_{l,s}$-axis is divided up into intervals ${\delta}\rho_{l,s}$.
For each $\rho_{l,s}$ the respective inverse image area in the
$\eta^{\perp}\zeta^{\perp}$-plane is calculated for the values
$w_s(\eta^{\perp},\zeta^{\perp})$ belonging to $\rho_{l,s}$ and
${\delta}\rho_{l,s}$.  The total area is proportional to
$H(\rho_{l,s})$.  The chain intersection function
$w_s(\eta^{\perp},\zeta^{\perp})$ is piecewise linear due to the
flatness of the surface rhombs of the triacontahedron.  It decreases
monotonously in radial direction from the center of the interface
$\phi_{Tria,s}$ to the boundaries.  

\begin{figure}
\caption{Chain structures in the ideal tiling. The histograms of chain
  densities $H(\rho_{l,s})$ are given.\label{chainid}}
\begin{center}
\vspace*{1cm}
\includegraphics[width=6.5cm,angle=0]{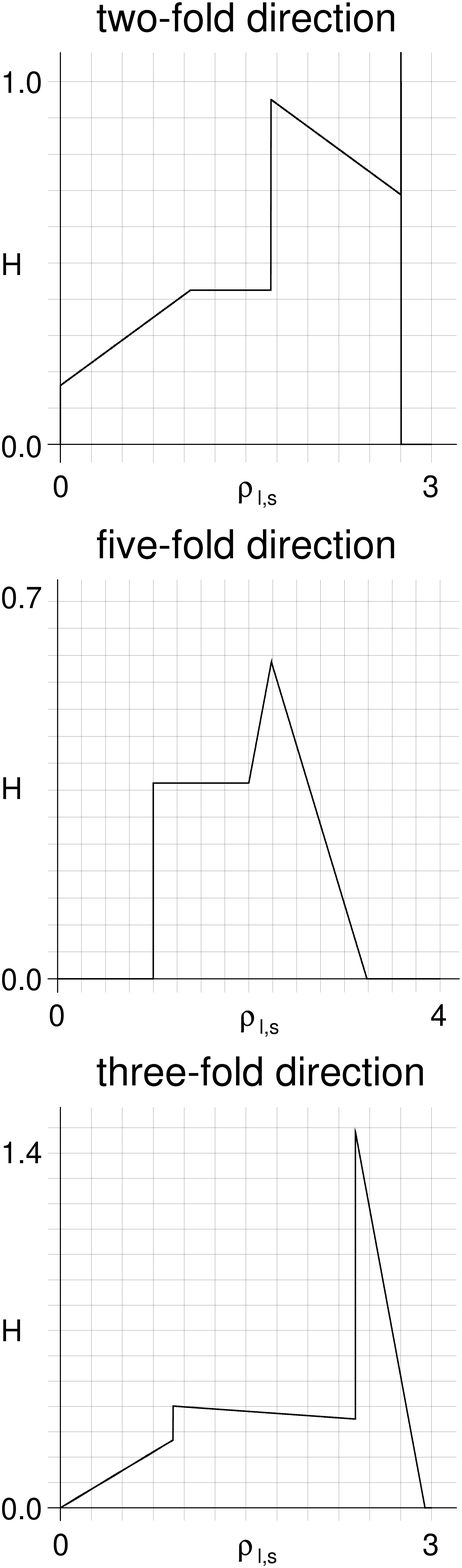}
\end{center}
\end{figure}

The frequency function may be derived analytically in the following
way: The origins of the coordinate $\xi^{\perp}$ parallel to the
symmetry axis ${\hat s}$ and of the coordinates $\eta^{\perp}$ and
$\zeta^{\perp}$ in the plane perpenticular to ${\hat s}$ are chosen in
such a way that it 
coincides with the projection of the center of the triacontahedron.
Then we compute the area of the interface $F(\xi^{\perp})$ of the
triacontahedron with the same $\xi^{\perp}$ as a function of
$\xi^{\perp}$ for the half space $\xi^{\perp} \geq 0$.  Due to the
convexity of the triacontahedron the function $F(\xi^{\perp}$)
decreases monotonously. From the inverse of the derivative of
$F(\xi^{\perp})$ (${\delta}\rho_{l,s}$$\to$0) one obtains a function
which is piecewise linear, has discontinuities and can be identified
with the density function $H(\rho_{l,s})$ (Fig.~\ref{chainid}).

\section{Energy measures}
\label{energy}

\subsection{Models of stability} \label{ucs}
Several models currently exist to explain the stability of
quasicrystals.  In the deterministic energy model the internal energy
$U$ represents the thermodynamically stabilizing factor.  Microscopic
forces lead to matching rules~\cite{debruijn} or overlapping cluster
energies~\cite{gummelt,heong} that favour an ideal quasicrystalline
tiling. In the non-deterministic entropic model stabilized by the
entropy $S$ no matching rules exist but the cells of the tiling do not
leave any gaps and do not overlap. This is the random tiling model. In
between is the semi-entropic model which is described by a free energy
with contributions from internal energy and entropy:
\begin{equation}
F(T)=U(T)-TS(T)\quad.
\end{equation}
The purely energetic model can be regarded as a low-temperature limit,
the purely entropic model as a high-temperature limit of the
semi-entropic model.

\subsection{Properties of energy measures}\label{enemeas} 

The ideal quasilattice without any violation of matching rules
represents the ground state of the energy model, taken at $T$=0.
At finite temperatures thermally activated flips exist, mediating the
transition between neighbouring states. The transition probability
is given by the Boltzmann factor, which depends on the energy measure
chosen. 

In this work we deal with the canonical random tiling ensembles. All
configurations of an ensemble have the same volume and the same
number of vertices due to a constant average slope of the
de-Bruijn-hypersurface. The energy of the system is the
ensemble average. The pure entropic random tiling model, on the other
hand, is spezialized for microcanonical ensembles, since all
configurations have the same energy.  Starting from the occupation
distribution of the energy levels of a canonical random tiling
ensemble in the thermodynamic equilibrium we can calculate the
internal energy $U(T)$ as the ensemble average $U=<E>$ of the
instantaneous energy $E$. The specific heat $C_{V}(T)$ can be derived
from the variance of the occupied energy levels by the
fluctuation-dissipation theorem
\begin{displaymath}
<(\delta E)^2> = <E^2>-<E>^2 = k_B T^2 C_V =\frac{\partial U}{\partial
T}\quad. 
\end{displaymath}

The temperature-dependent internal energy and the specific heat are
governed by the intrinsic properties of the system and the selected
energy measure.  The temperature dependence of the entropy density
$s$($T$) (precisely: entropy per quasilattice site\footnote{Large
  letters indicate total quantities, small letters denote a quantity
  per vertex. For periodic approximants with rhombohedral cells the
  number of cells and vertices is identical.}) is given by the
thermodynamical integration of 
$c_V$:
\begin{equation}
s(T)=s_V(T) +\int\limits^T_0
dT^{\prime}\frac{c_V(T^{\prime})}{T^{\prime}}+s_0\quad.
\end{equation}

$s_V$($T$) is the entropy contribution of a change of volume. It
vanishes for canonical random tilings. $c_V$ is the isochoric specific
heat and $s_0$:=$s$($T$=0) the ground state entropy. At
$T$$\rightarrow$$\infty$ we are in the limit of the pure random tiling
model and we get the configuration entropy $s$($T$=$\infty$)
=:$s_{\infty}$. The temperature variation of the specific heat depends
on the energy measure, but not $s_{\infty}$. If one is interested only
in the configuration entropy $s_{\infty}$, then the specific choice of
the energy measure has no physical relevance if the first three of the
following four conditions are fulfilled:
\begin{enumerate}
\item The energy of any configuration is unique.
\item The quasiperiodic reference tiling is a ground state.
\item The energy measure is limited from above.
\item The entropy of the ground state vanishes or is easy to
calculate.
\end{enumerate}
The first condition requires an exact law to determine the energy of
any microstate. The energy of a fixed microstate must not depend on
how it is generated. Two configurations that differ only by a rigid
translation have to be energetically identical.

Especially global energy measures are endangered to violate the second
condition and to render the quasiperiodic reference state unstable at
$T=0$. The third condition is a requirement for the integrability of
$\frac{\displaystyle c_V(T)}{\displaystyle T}$ 
as a function of $T$. If the energy measure is not
limited, uncontrolled fluctuations of the energy in the
high-temperature limit may exist. If they vary stronger than quadratic
with $T$ then
\begin{displaymath}
\frac{<(\delta E)^2>}{T^2}
\end{displaymath}
diverges and with it $c_V$ as $T$$\rightarrow$$\infty$.\\

The last condition need not be met but is recommended for practical
purposes. The ground state entropy $s_0$ represents the number of
states energetically equivalent to the quasiperiodic ground state.
Since it is often complicated to calculate the ground state entropy it
is useful to choose an energy measure where $s_0$ vanishes or is easy
to deduce. An example of a complicated case of non-vanishing
contribution $s_0$ has been presented by Baake and
Joseph~\cite{baake90} for a two-dimensional octagonal quasicrystal.
The 
locally defined energy measure is 0 for the canonical vertices but
positive for forbidden vertices and depends on the vertex type.
Unfortunately there exist configurations which are not quasiperiodic,
although they do not exhibit forbidden vertices. As a consequence 44\%
of the total entropy are contributed by the ground state.

An example for a globally defined energy measure with an easily
calculable ground state entropy is the ansatz of a harmonic oscillator
potential in the orthogonal space as applied in this work.  The energy is
the sum of the squared distances of the dual quasilattice
sites from their center of mass. Up to a factor $N$ it is equal to the
variance in the orthogonal space (see Sec.~\ref{variance}).  The
precise definition of the energy of the configuration $\alpha$ is
\begin{equation} \label{energymeasure}\label{eneq} 
E^{(\alpha)}=C
\left|
 \sum^N_{j=1}
    \left|
      {\bf y_j^{(\alpha)}} - \frac{1}{N}\sum^N_{i=1}{\bf y_i^{(\alpha)}}
    \right| ^2
- \sum^N_{j=1}
    \left|
      {\bf y_j^{(0)}} - \frac{1}{N}\sum^N_{i=1}{\bf y_i^{(0)}}
    \right| ^2
\right|\quad.
\end{equation}
The index 0 denotes the ideal reference configuration,
$C$=$\frac{\displaystyle 250}{\displaystyle 61}$ is an arbitrary
normalization constant.  

This energy measure is called the {\bf harmonic energy measure}. The
variance for the ideal reference configuration 0 is smaller than the
variances for the overwhelming majority of the random tiling
configurations $\alpha$. But there is a tiny minority of
configurations with a variance smaller than the value of the ideal
tiling. Their atomic hypersurfaces are closer to a sphere than the
triacontahedron. Such configurations may play a r\^{o}le at very low
temperatures. To avoid energies less than the energy of the ideal
tiling we have taken the absolute value in Eq.~\ref{energymeasure}.
For zero global phason strain in an ideal tiling this energy measure
is not degenerate. But for periodic approximants the $N$ possibilities
($N$: number of lattice points) to chose the origin of the unit cell
yield a ground state entropy 
$s_0=\frac{\displaystyle \ln N}{\displaystyle N}$ per lattice point.
It is obvious that $s_0$ vanishes in the thermodynamic limit.
Strandburg~\cite{strandburg91} has introduced a similar energy
measure. But it was taken relative to a fixed point in the orthogonal
space and not relative to the center of mass and therefore does not
fulfil the criterium of finiteness of the energy measure (third
condition).  Without fixed boundaries of the system
the whole distribution of the dual quasilattice
points in the orthogonal space may drift
and therefore yield a systematic contribution to the energy.

Other global energy measures (not limited!) which we have used on a
trial basis are the {\bf cubic energy measure}
\begin{displaymath}
E^{(\alpha)}=\sum^{N}_{j=1} \left|{\bf y}_j^{(\alpha)}-{\bf M}_{Tria}
\right|^3
\end{displaymath}
and the {\bf quadratic energy measure}
\begin{displaymath}
E^{(\alpha)}=\sum^{N}_{j=1} \left|{\bf y}_j^{(\alpha)}-{\bf M}_{Tria}
\right|^2
\end{displaymath}
(the latter one has been introduced by
Strandburg~\cite{strandburg91}). ${\bf M}_{Tria}$ denotes the fixed
center of mass of the triacontahedron in the orthogonal space.

An example of a locally defined energy measure is the one that counts
the violations of the alternation condition. This condition requires
that along a worm two rhombohedra of the same type and orientation do
not occur subsequently.  This energy measure was used by Dotera and
Steinhardt~\cite{dotera94} and G\"ahler~\cite{gaehler95}. Other
examples are the ``simple energy model'' which assigns the same energy
to each forbidden verte, the ``cluster energy model'' which
maximizes the frequency of certain favourite vertices and the
``T\"ubinger mean-field model''. All three where used
by Joseph in~\cite{joseph95}.

\subsection{Pair interactions} 

All energy measures discussed in Sec.~\ref{enemeas} contain a
certain kind of arbitrariness since they do not depend on interatomic
interactions but on the assignment of an increased energy value to
randomized lattice configurations. Here we present test interactions
which are motivated by pairwise interactions between atoms sitting on
certain sites of the quasilattice (called ``decoration''). We test
whether the energetic equidistribution postulated in the random tiling
model is justificable.

The potential interactions used are the 12-6-Lennard-Jones potential
modified by a cutoff function to guaranty a smooth behaviour at the
cutoff radius ($r_c = 4$ in units of rhombohedra edges). Another
model is the potential of mean force
\begin{displaymath}
v(r)=-\ln(g(r)+\kappa),\quad \kappa> 0.
\end{displaymath}

$g(r)$ is the radial density function and $\kappa$ an arbitrary
parameter which screens the singularity caused by the zeros of $g(r)$
($\kappa=100$ in this work). To arrive at a potential which does not
consist of purely delta-like minima due to the delta-like maxima of
$g(r)$ of an ideal quasicrystal one usually broadens the maxima with a
gaussian. This yields an additional free parameter, the width (if all
the maxima are broadened by the same function).

The results of the test are presented in Sec.\ \ref{resrdf}. It turns
out that different random tiling samples are indeed equivalent with
respect to the pair interactions, since the radial distribution
function is nearly unchanged for the short distances (where the
contibution to the potential energy would be strong), and at larger
distances the pair potential is zero due to the cutoff distance.

%

\section{Construction of random tilings}
\label{construct}

\subsection{Monte-Carlo method} 

Simpleton flips directed in their frequency by the Monte-Carlo method
allow the generation of equilibrium configurations at a given
temperature. The randomization of the quasilattice is
realized by a sequence of such flips which usually are associated
with a change in energy. The flips play the role of elementary
transition paths between different states of a system in contact with
a heat bath.

We are assuming that the method is ergodic~\cite{gaehleryy}. Although
this has not yet been proven, there is no hint up to now of the
opposite.  The method has already been used extensively to generate
and study equilibrium random tiling quasicrystals. It is the only one
known that produces equilibrium ensembles.

A simpleton flip does not generate or destroy tiles, it only
rearranges them. Therefore the frequency of both types of rhombohedra
in the tiling is remaining constant. The average orientation of the
hypersurface and the global phason strain are not changed.

During the Monte-Carlo simulation lattice points are selected at
random. If a lattice point belonging to the simpleton vertex is hit,
the energy of the original and the flipped configuration are compared.
If the energy decreases, the flip is always carried out. If it
increases, it is performed with a probability of $\exp(-\frac{\Delta
  E}{k_{B} T})$.

During the flips the lattice points change their positions and the
vertices their frequency. The simpleton vertex itself, for example,
occurs with a frequency of 23.61\% in the ideal tiling. This value
decreases to a temperature dependent equilibrium value during the
initial phase of a simulation. At $T=\infty$ the equilibrium frequency
of the simpleton vertex is about 17.5\%.

\subsection{Boundary conditions} 

Quasicrystals do not permit periodic boundary conditions. Hence one
has to work with a finite patch of a quasicrystal. Then there are
two possibilities to deal with the surface: either keep it fixed (fixed
boundaries) or identify opposite sides (periodic approximants). 

In the case of fixed boundaries the surface lattice points are not
allowed to move.  In approximants all lattice points are mobile.
Therefore the configuration entropy per lattice point with fixed
boundaries is smaller than the entropy of approximants.  The
configuration entropy of the approximant, on the other hand, differs
only slightly from the value of the ideal tiling. The frequency of
vertices also is not much different between an ideal quasicrystal
and an approximant tiling. The frequency of the simpleton, however, in
patches with fixed boundaries may grow to up to 30\% of that of the
bulk vertices. For a discussion of the entropy dependence on boundary
conditions see~\cite{joseph96}.

Periodic approximants show matching rule violations already in the
ground state. They exhibit a periodic superstructure and for very
small samples a remarkable deviation of the vertex frequencies. The
intrinsic global phason strain changes the flip diffusion properties
at low temperatures as the ground state phasons generate zero energy
modes. The sheet magnetization (see Sec.~\ref{sheetmag}) in the case
of a cubic approximant is useful only for the three two-fold
directions parallel to the cubic cell axis. The acceptance region is
no longer densely covered but has a lattice structure. This fact
causes problems in the averaging procedures described in
Sec.~\ref{channeling}.

\subsection{Addition method} 

A new method to generate random tilings (especially non-equilibrium
configurations) is the successive addition of tiles to the tube and
slab structures within a finite region.

For this method fixed boundaries are used. We start with a patch of a
tiling with a fixed (possibly zero) global phason strain.  This can be
e.g.\ a piece of an ideal tiling or a single unit cell of a periodic
approximant. All interior quasilattice points are removed. Then the volume is
refilled in a systematic way without gaps and overlaps of the tiles.
Since we start with an existing patch (microstate), it is certain that
tilings of this surface must exist. The hollow volume represents the
macrostate, the different possibilities to fill it are the
microstates.

\begin{figure}
\begin{center}
\caption{Intermediate state in the addition process. The star
  indicates the tiling vector directions. The double arrow indicates the
  direction which has already been filled. The small arrows
  indicate the correlations to the direction which will be
  filled next.\label{addiproc}}
\vspace*{1cm}
\includegraphics[width=14cm,angle=0]{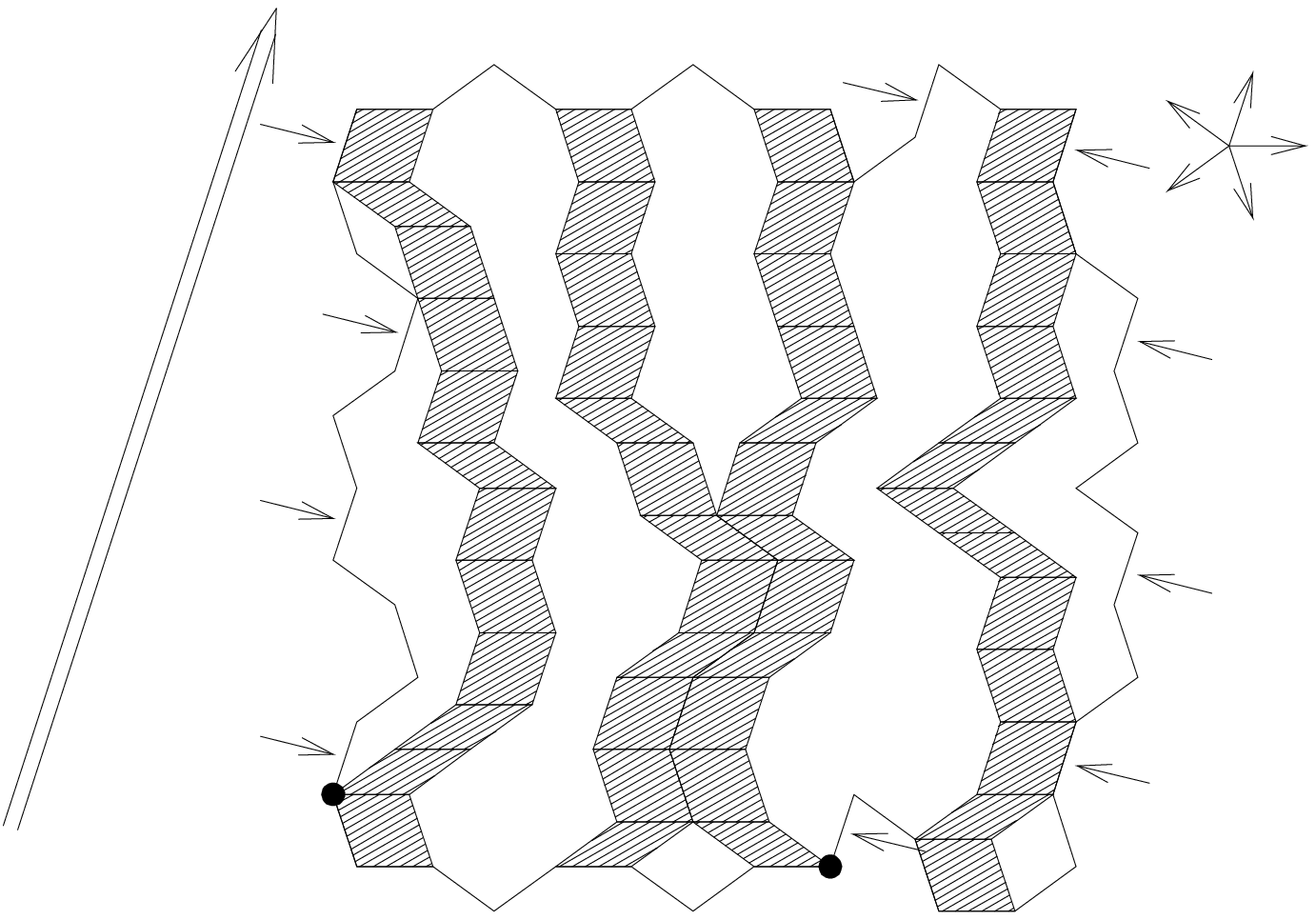}
\end{center}
\end{figure}

The algorithm is started by selecting one five-fold direction
(Fig.~\ref{addiproc}).  First 
the ensemble of parallel slabs perpendicular to this direction has to
be filled with rhombohedra. The borders of the slabs are given by a
surface consisting of rhombi parallel to the five-fold axis chosen. A
slab appears as a pentagonal Penrose rhombus pattern in projection.

The surface of the hollow volume and the correlation between the slabs
dictate the addition conditions. A violation sooner or later leads to gaps or
overlaps, and the configuration has to be abandoned. 

We will now shortly sketch how the method works. The procedure can be
broken up into two parts: filling the two-dimensional slabs, which is
equivalent to performing the construction for generalized Penrose
tilings, and then filling the space between the slabs. 

First of all we will explain how to proceed in two dimensions.  In the
Penrose tiling there are five sets of one-dimensional rows of rhombi
with one type of edge in common. The average direction of the sets is
perpendicular to the tiling vector directions. In the hollow volume we
start with one of the directions and generate the rows one by one. The
fixed surface determines the starting and terminating position of the
rows (by the edges parallel to the selected tiling vector). The
difference vector (in higher dimensional coordinates) between initial
and final point determines the number of tiles of each type in that
row uniquely (by the coordinate components). We are only free to
choose the sequential arrangement of the tiles.  There are, however,
restrictions: The second pair of edges of each rhombus must also fit
into the configuration.  This means that the difference vector between
the rhombus and the respective edge on the surface of a neighbouring
row must be fillable with rhombi. Otherwise, the configuration has to
be abolished. After all the rows of one direction have been created,
the next direction is being dealt with until the whole patch is
filled. The last direction is redundant, since the whole slab
is filled as soon as the second-to-last direction is worked out.

Up to a certain degree the process is running analogously in the three
dimensional case. The slabs are being filled in the same way as for
twodimensional Penrose-like rhombus tilings.  We continue with the
next slab perpendicular to the selected five-fold direction and
generate slab by slab along this direction. Now we have to check the
difference vectors of two pairs of edges to make sure that the space
in between can be filled.  But one must be careful as there are row
directions crooked by oriented with respect to the row directions within
each slab.  For example, the row direction perpendicular to the
rhombus spanned by the vectors ${\mathbf t}_2$ and ${\mathbf t}_3$ are
crooked with respect to the corresponding direction of the vectors
${\mathbf t}_0$ and ${\mathbf t}_1$. To explain how to overcome this
problem, however, takes more space than is available here. Interested
readers are 
referred to~\cite{ebinger96}.  After all slabs of one direction are
constructed the other directions are attacked until the whole volume
is filled.  Tilings generated by this method are called ``addition
tilings''.

\subsection{Comparison of the methods} 

\begin{figure}
\begin{center}
\caption{Comparison between different tilings: a) perfect
  quasicrystal, b) generated by flip process, c) generated by addition
  process. The corresponding acceptance domains are labeled by d), e)
  and f). \label{addproc}}
\vspace*{1cm}
\includegraphics[width=14cm,angle=0]{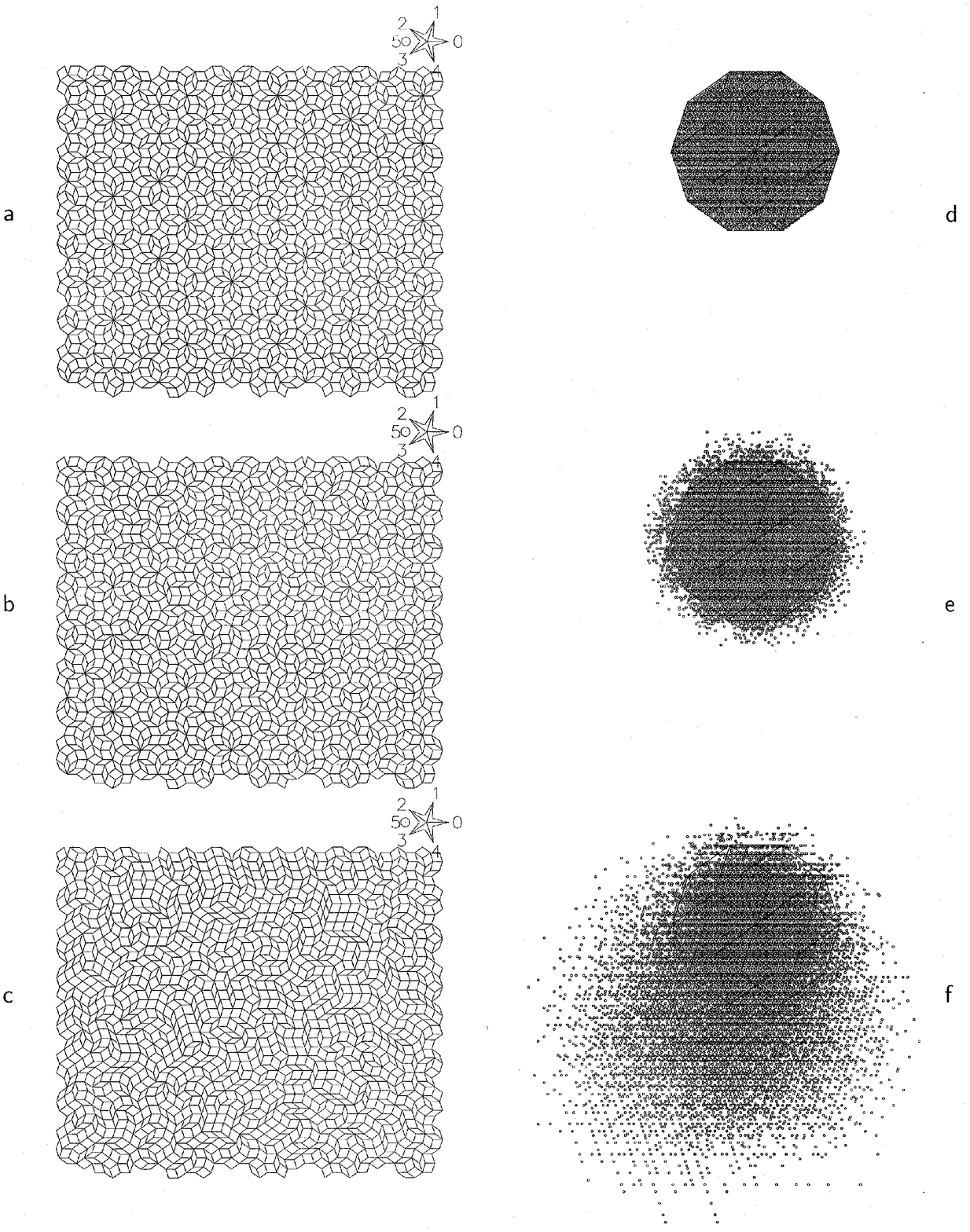}
\end{center}
\end{figure}

The acceptance domain for the ideal icosahedral rhombohedral tiling is
a triacontahedron (Fig.~\ref{addproc}). In an equilibrium configuration
generated by the 
simpleton flip method it is smeared out to an almost Gaussian
distribution. Differences in the physical space are relatively small.
There are some matching rule violations like pairs of adjacent rhombi of
the same type, but the structure does not appear to be massively
disordered. The phason fluctuations of the configurations are limited
according to~\cite{tang90}.

In a configuration generated by the addition method the deviation from
quasiperiodicity is much stronger (Fig.~\ref{addproc}): There are
microcrystalline 
inclusions which indicate large-wavelength fluctuations in the
higher-dimensional space. Many twin, domainlike or grain boundaries
occur. The deviation from the ideal acceptance domain is strong and
anisotropic. Configurations far from equilibrium are easy to
construct.

The efficiency of a specific method depends on its ergodicity
properties. In a configuration generated by the addition method with
microcrystalline inclusions there are few possibilities for flips.
Very often the few flippable lattice points only jump back and forth.
There are bottle necks which prevent or strongly inhibit the structure
from moving through phase space.

Furthermore a configuration depends on its history. To generate
independent configurations one has to wait until the autocorrelation
is dropped off sufficiently. At high enough temperatures about 50 
Monte-Carlo steps
are sufficient. Fixed boundaries may increase the correlation times
drastically. 

The anisotropic character of the configurations generated by the
addition method is due to the fact that the slabs are filled
sequentially. The deviation of the center of mass of the point
distribution from the original acceptance domain depends on the
sequence of construction of the slab and tube directions. On the other
hand, it is only the sequential treatment that permits the
tractability of the method.

In the addition method the probabilities for
adding a certain rhombohedron at a certain place have not been taken into
account. Therefore, the configurations generated by this 
method in a selected hollow volume can not be considered as the sample
survey of a equilibrium ensemble. To calculate the probabilities would
be impossible since the number of realizations is prohibitively high.

The addition method allows the generation of configurations that are
never reached in acceptable times by the simpleton flip process. The
method could be used to generate random tilings with constraints.

Experimentally, a number of decagonal quasicrystals have been found
that maybe could be described by configurations generated by the addition
method~\cite{hiraga91a,he91,hiraga91b,josephxx}. The distribution of
the quasilattice points in the orthogonal space is not isotropic as in the
samples produced by the simpleton method but strongly anisotropic.

\section{Random tiling transition and structure functions} 
\label{random}

Signals for a random tiling transition can be obtained in a number of
ways. The thermodynamic functions $u$, $c_V$ and $s$ may show a
characteristic behaviour. For example a plot of the specific heat
$c_V$ versus temperature for different sample 
sizes (finite-size scaling) may indicate the possibility of a
random tiling transition. The specific heat $c_V$ should diverge in
case of a second order phase transition.

For any desired temperature a histogram of frequency of the energy
values can be calculated during the simulation. The histogram for
infinite temperature yields an approximation for the density of 
energy states since at infinite temperatures all energy levels should
be occupied with equal probability. Therefore it may be used to explain
the behaviour of the specific heat as a function of $T$ (see also
Sec.\ \ref{ieshsn}).

In addition to the specific characterization criteria for quasicrystals
described in Sec.~\ref{character} there are more general functions
that can indicate a phase transition. We have applied the following:

\subsubsection{Self diffusion coefficient}
The average squared difference $\left <({\bf r}(t) - {\bf r}(0))^2
\right>$ of the coordinates of all quasilattice points from their
initial position is expected to depend linearly on time at long enough
simulation times. The slope of the function is proportional to the
self-diffusion coefficient $D$. A change of the slope may indicate a
phase transition which involves the change of energy barriers for example.

\subsubsection{The Binder order parameter} 

The Binder order parameter~\cite{binder} $B^{(N)}(T)$ for a given
sample size $N$ and temperature $T$ is defined by 
\begin{displaymath}
B^{(N)}(T):=1- \frac{M_4^{(N)}(T)}{3\left(M_2^{(N)}(T)\right)^2}
\end{displaymath}
where $M_2^{(N)}$($T$) and $M_4^{(N)}$($T$) are the second and fourth moment 
\begin{displaymath}
M_k^{(N)}(T)=\int d\mu \left(p(\mu)^{(N)}(T)\right)\mu^k
\end{displaymath}
of the probability $p(\mu)$ of a microstate of the sheet magnetization
$\mu$  belonging to a 
macrostate with sheet magnetization $M$. The Binder order parameter
$B^{(N)}(T)$ is plotted as a function of $T$ and yields a set of
curves parametrized by $N$. A unique intersection point of
the curves points to a second order phase transition
and yields the transition temperature. If there is
no clear intersection point only an interval for a possible phase
transition can be determined.

\subsection{Structure functions}\label{structure}

\subsubsection{Radial distribution function $G(r)$} 

The radial distribution function $G(r)$ of lattice points or atoms (in
case of a decorated quasilattice) consists of discrete delta-type
maxima and is related to the radial density function $g(r)$ by
$G(r)=4\pi r^{2}g(r)$. The Fourier transform of the radially averaged $G(r)$ is the
radial structure factor $I(k)$. The potential energy $u$ of an
quasicrystal can be directly calculated from $G(r)$ if there are only
pair interactions $v(r)$:
\begin{equation}\label{pairpot}
u=\sum_{r=0}^{R} G(r)v(r)
\end{equation}
where $R$ is the cutoff radius of the potential.
If there are different types of atoms, then additional sums over
the different types of interaction pairs must be taken.

\subsubsection{Structure factor intensity $I(k)$} 

Flips of the tiles modify the occupation density of atomic
chains and atomic layers. Diffraction diagrams are dominated by the
long-range translational and orientational ordering of
quasicrystals. Therefore an alterations of the quasiperiodicity changes
the width and intensity of reflexions.  

Phason fluctuations furthermore generate diffuse scattering. 
Due to the limitation of the phason fluctuations~\cite{tang90}
the maximal intensities are barely 
influenced. Diffuse scattering is governed by the
phason-elastic constants.

\subsubsection{Diffraction diagrams}
Diffraction diagrams for random tilings were calculated for
randomly selected sample configurations filed during the simulation
process at a given temperature $T$ and size $N$ and
averaged over the whole set chosen. The fluctuations between the
selected configurations turned out to be small
at a given $T$ and $N$. 

\section{Simulation process and models studied} 

The internal energy $u$, the specific heat $c_V$, the entropy $s$ as
well as the sheet magnetization $M$, the corresponding susceptibility
$\chi$ and the self-diffusion coefficient $D$, all dependent on the
temperature $T$, are thermodynamical state variables of a
random tiling ensemble.  The structure of a single tiling, but also
the whole ensemble can be characterized by structural functions. In
this work we have used the diffraction-intensity $I(k)$ (without
paying attention to diffuse line shapes), the relative frequencies of
vertex environments, the axial and planar structures, the variance
$\Omega$ (phason fluctuations) and the radial distribution function
$G(r)$.  The quantities $u$, $M$, $\Omega$ and $\left< ({\bf r}(t) -
  {\bf r}(0))^2 \right>$ \footnote{The simulation time is set to zero
  after thermalization.} are sampled by registring their values at regular
intervals of length $\Delta t$ during the simulation.
From the collected data all above mentioned state variables are
calculated.  Data of microconfigurations are stored every 1000$\cdot$$\Delta
t$ Monte-Carlo step. The results may deviate from the thermodynamical
limit due to finite-size-effects. If the sample size grows, the values
should converge to the thermodynamic limit.

An initial configuration (e.g. an ideal Ammann-Kramer-Penrose
approximant or an addition configuration) was equilibrated over a
typical thermalization time of $10^3$ to $10^4$ steps before the real
simulation was started. The thermalization time was chosen according
to the saturation behavior of the energy, for example.  At low
temperatures longer equilibration and simulation times and sampling
intervals were used.  The length $\Delta t$ was checked by the decay
of the autocorrelation functions.  For high temperatures $\Delta t$
was set typically about 50 Monte-Carlo moves per lattice point,
while for lower temperatures 200 moves were taken due to temporal
correlations.  The entire duration of a simulation run at given
temperature typically has been about 25000$\cdot$$\Delta t$ to
30000$\cdot$$\Delta t$ simulation steps.

\subsection{System sizes and interaction types}

The following systems have been studied in our simulations:

\begin{enumerate}
\item A finite patch of a quasiperiodic lattice with 4403 lattice
  points (3507 cells), fixed boundaries and the cubic energy measure.
  Starting configuration has been the ideal quasilattice.
\item A cubic approximant with 2440 lattice points, periodic
  boundaries and the quadratic energy measure.  Starting
  configuration was the ideal approximant.
\item Two five-fold approximants with 890 and 1440 lattice points,
  periodic boundaries and the harmonic energy measure.  Starting
  configuration also has been the ideal approximant.
\item Five cubic approximants with 136/712 (n=3), 576/3016 (n=4),
  2440/12776 (n=5), 10336/54120 (n=6) and 43784/229256 (n=7) lattice
  points/atoms in the binary decoration model\footnote{In the binary
  model each vertex and edge center is decorated with a small atom,
  and the long diagonal of the prolate rhombohedron is decorated with
  two large atoms}~\cite{henley86} ($n$ is
the generation), 
  periodic boundaries and the harmonic energy measure. In this
  category we used ideal approximants as well as addition tilings as
  starting configurations. Here we also calculated the self-diffusion
  coefficient $D(T)$, the magnetization $M(T)$ and the susceptibility
  $\chi(T)$.  The results are independent of the starting condition in
  the case of periodic boundaries: Ideal quasicrystals and addition
  tilings yield the same ensemble averages\footnote{which is not true
    in the case of fixed boundaries}, proving that the system had
  been well equilibrated before data collection.
\end{enumerate}

\section{Results}
\label{results}

\subsection{Direct calculation of the configurational entropy}

For small clusters of rhombohedra with fixed surface it is possible to
calculate the configurational entropy $s$ directly, which we have
performed for the five smallest polyhedra with triacontahedral shape
and edge lengths of one or two. The results are presented in
Tab.~\ref{tabent}.

The clusters are too small to permit a reasonable extrapolation to the
thermodynamic limit. The value of the entropy is also too small due to
the fixed boundaries.

\subsection{Thermodynamic functions}

\subsubsection{Internal energy, specific heat, entropy}\label{ieshsn} 

In sample one the ground state entropy vanishes. The internal energy
grows with $T$ and saturates at high $T$. Due the fixed boundaries the
saturation limit is finite although the energy measure is unlimited.
The specific heat grows until $T\approx 1$. Then it decreases and goes
to zero at high $T$ i.~e.\ we observe a \textsc{Schottky} anomaly as
is typical for two level systems. The entropy at infinite temperature
approaches $s_{\infty} \approx 0.1194 \pm 0.015$.

Sample two behaves similar, but the saturation of the internal energy
for $T > 10$ is not so clearly visible and may indeed not occur since
the energy measure is not limited and periodic boundaries were
applied in this case. The decay of the specific heat is slower which
is also a consequence of the energy measure. The entropy at infinite
temperature approaches $s_{\infty} \approx 0.2621\pm 0.015$.

\begin{figure}
\caption{Internal energy. Sizes are indicated by the number of 
lattice points.\label{intene}}
\begin{center}
\includegraphics[width=9cm,angle=270]{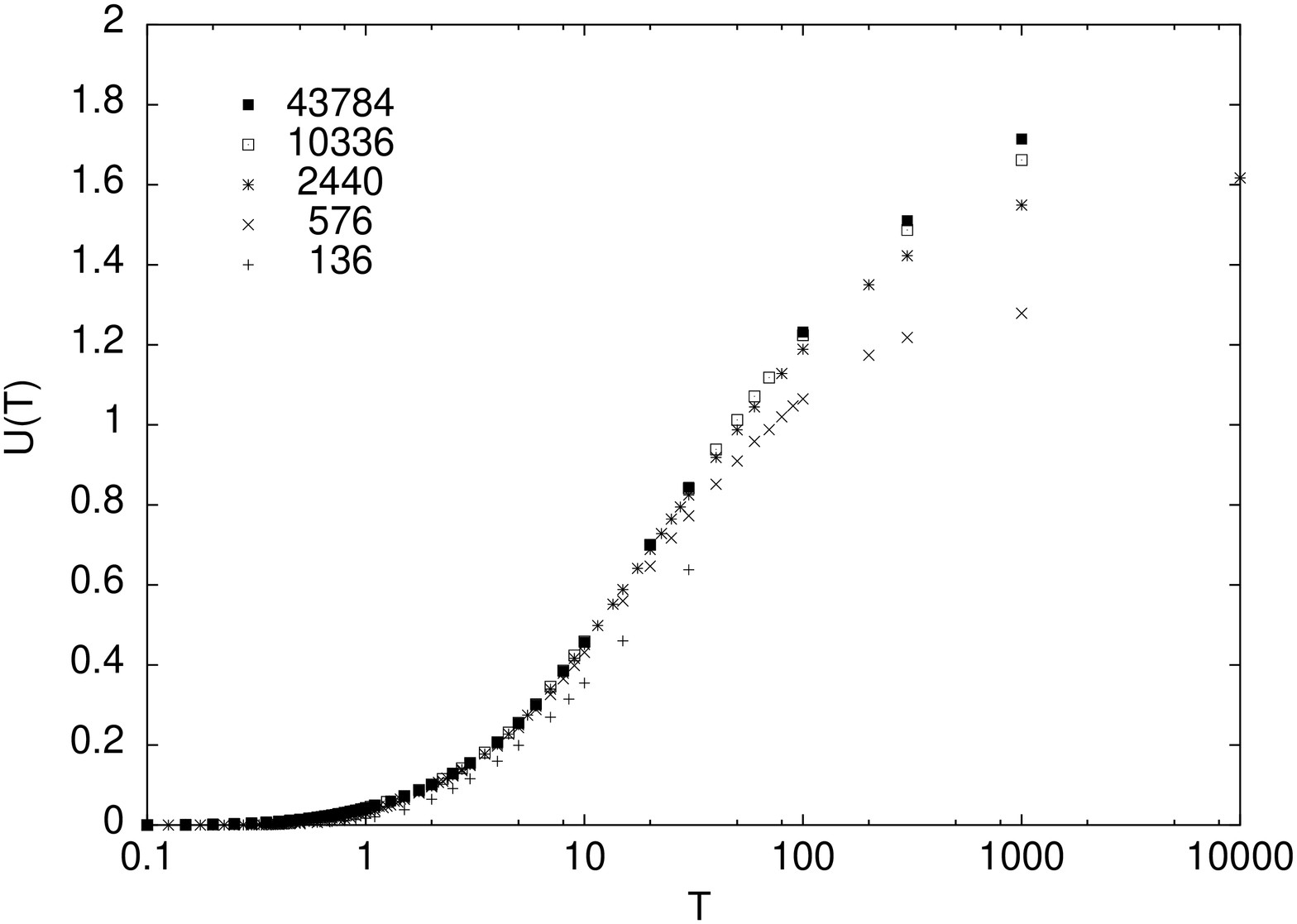}
\end{center}
\end{figure}

\begin{figure}
\caption{Specific heat. Sizes are indicated by the number of 
lattice points.\label{specheat}}
\begin{center}
\includegraphics[width=9cm,angle=270]{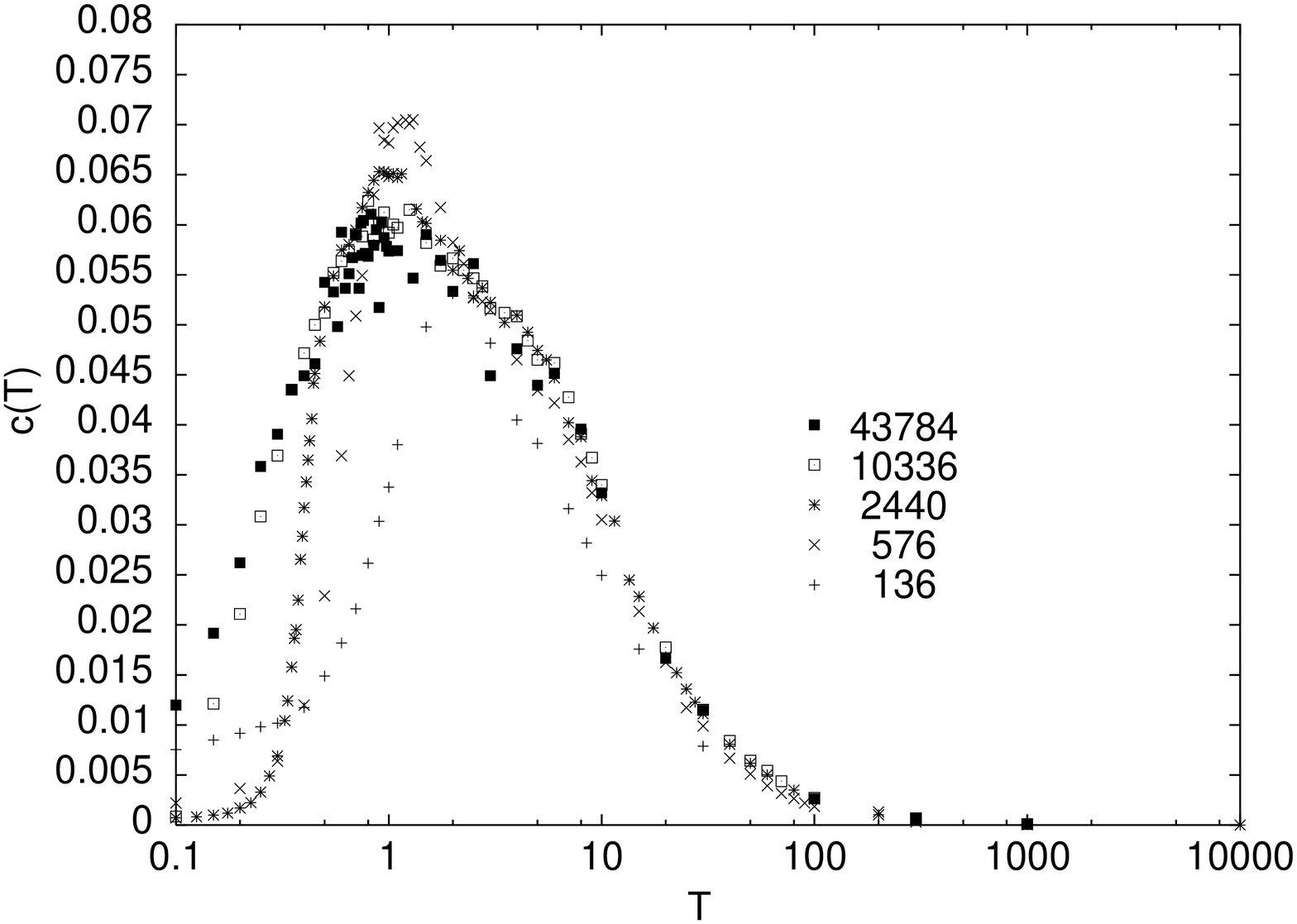}
\end{center}
\end{figure}

\begin{figure}
\caption{Configurational entropy. Sizes are indicated by the number of
lattice points.\label{confent}}
\begin{center}
\includegraphics[width=9cm,angle=270]{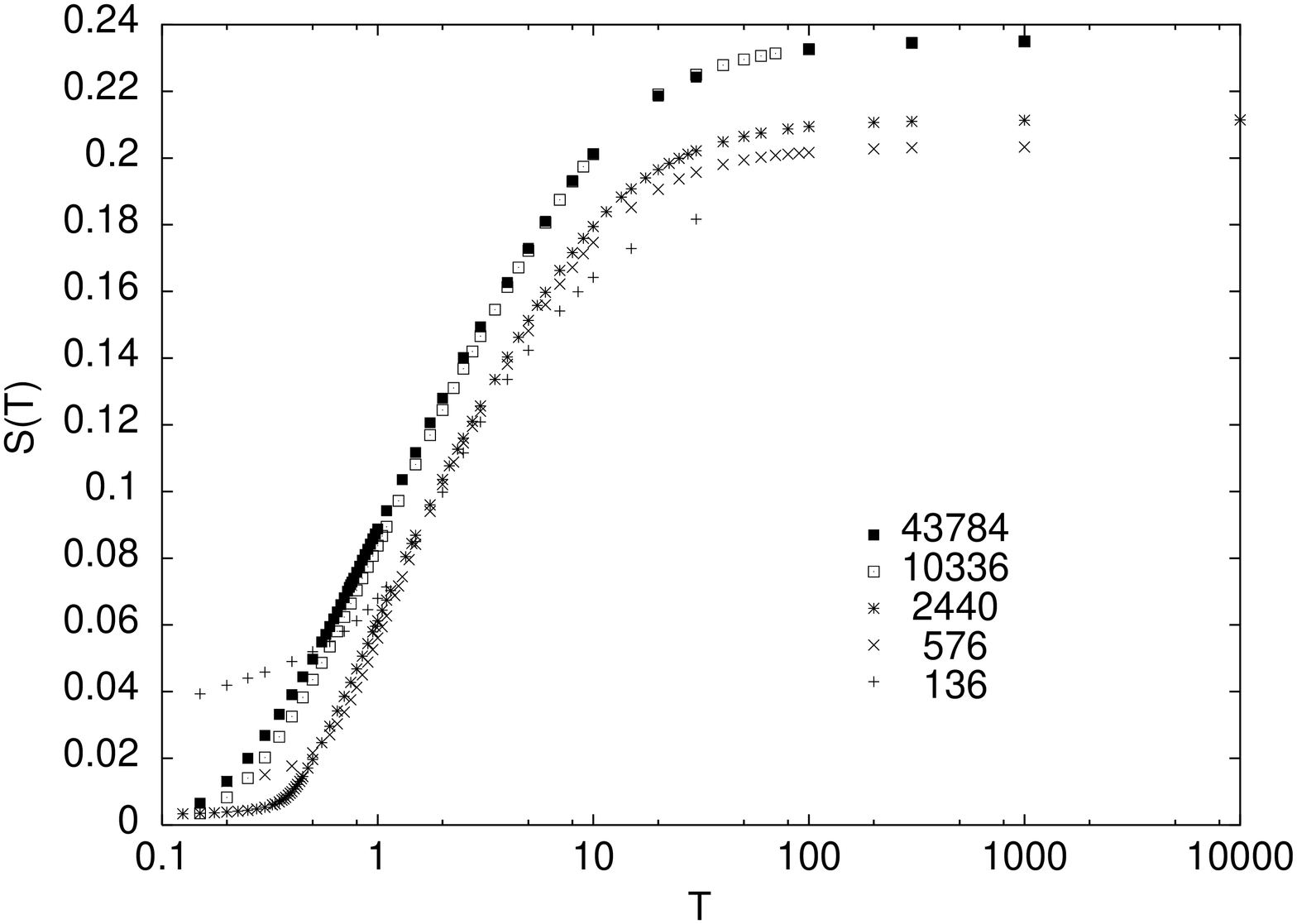}
\end{center}
\end{figure}

Samples three and four are again quite similar. The internal energy
grows monotonously, and a clear saturation becomes
visible (Fig.~\ref{intene}). The 
saturation value of $u$ depends on the size of the sample. The limit value
can be derived from limits of the variance: $\Omega$($T$=$\infty$) =
1.73 $\pm$ 0.01~\cite{tang90} leads to $u$ between 1.97 and 2.05.  A
\textsc{Schottky} anomaly in the $c_V$-Plot again is present but the
specific heat shows an additional bump above the
maximum (Fig.\ref{specheat}). Such a 
behaviour is known for few-level systems (for example three-level
systems) with sufficiently separated levels.  We have mapped the
distribution of the energy levels for $T$=$\infty$.  No indication of
discrete energy levels was found, only an asymmetry of the
distribution with a smaller slope at higher energies could be
observed.  There are other energy measures~\cite{gaehler95} which
exhibit no visible asymmetry in the energy distribution and no bump in
the specific heat. This, however, is in our opinion not a clear
explanation for the phenomenon. The value of the maximum of $c_{V}$ is
not significantly dependent on sample size. The increase in
$s_{\infty}$ is caused only by the growing width of maximum
(Fig.~\ref{confent}).  Entropy values at $T=\infty$ are listed in
Tab.~\ref{tab1}. Since no 
divergence occurs with increasing sample size, we have no hint
for a second order phase transition.
 
\subsubsection{Simpleton magnetization, simpleton susceptibility,
Binder order parameter} 

The temperature dependence of the energy fluctuations and the specific
heat do not give a clear indication for a phase transition. There is no
divergence of the maximum of $c_V$. 
The reason may be that the intrinsic divergence of the specific heat
with sample size, if any, is very weak. For further insight we
calculated the sheet magnetization $M$ and the susceptibility $\chi$
since the latter shows a much more pronounced divergence behaviour.

\begin{figure}
\caption{Magnetization. Sizes are indicated by the number of 
lattice points.\label{magnet}}
\begin{center}
\includegraphics[width=9cm,angle=270]{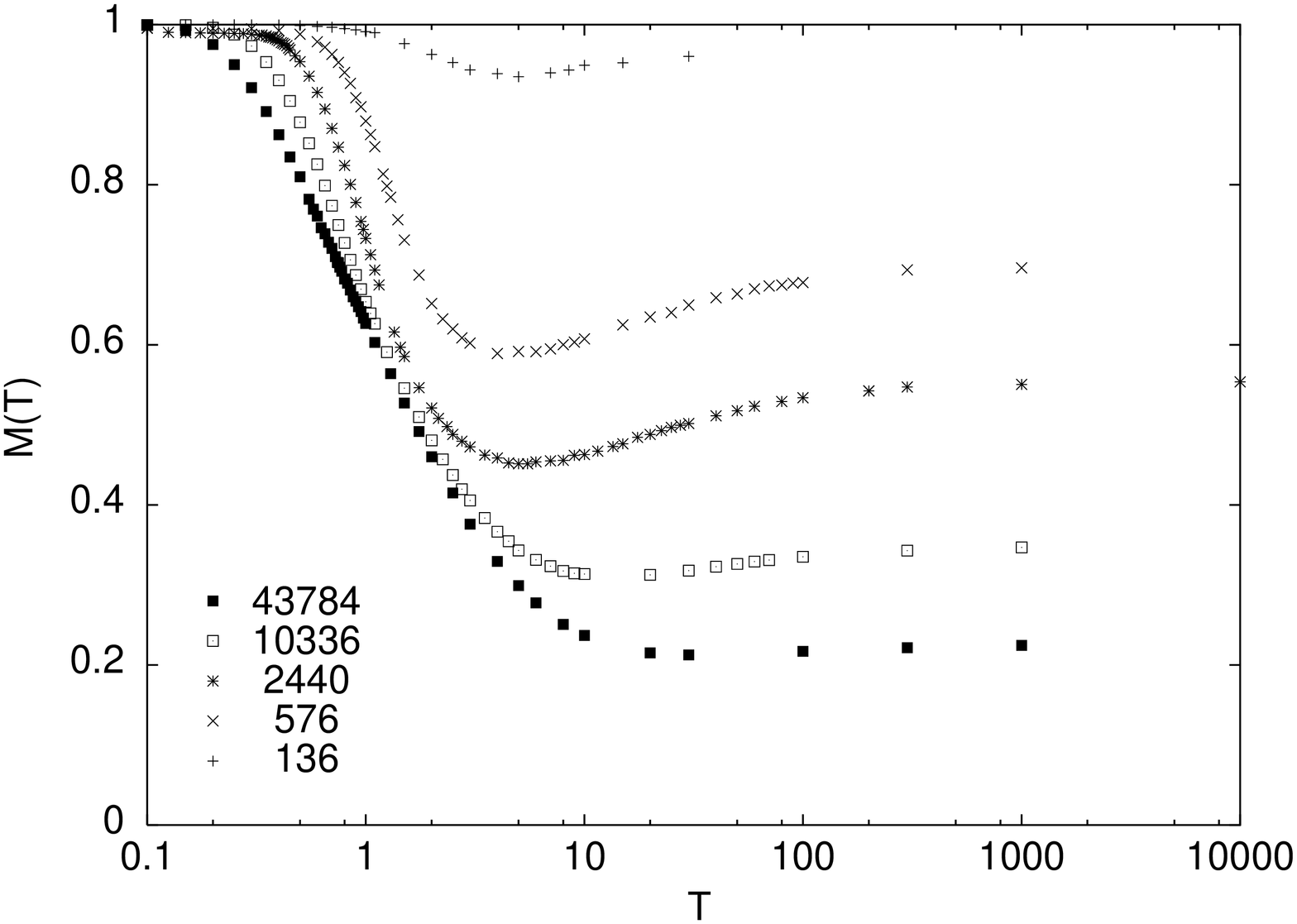}
\end{center}
\end{figure}

\begin{figure}
\caption{Susceptibility. Sizes are indicated by the number of
lattice points.\label{suscept}}
\begin{center}
\includegraphics[width=9cm,angle=270]{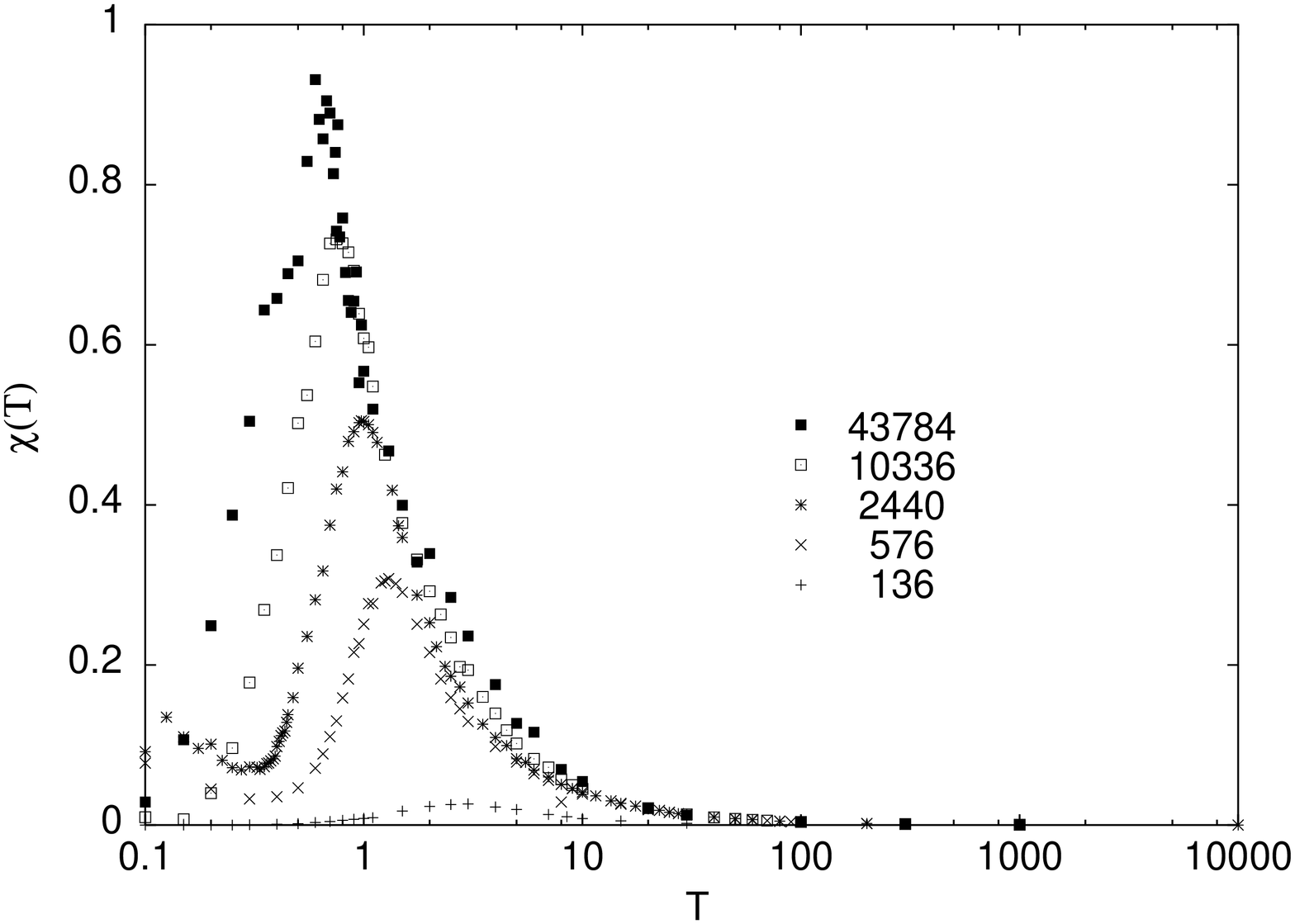}
\end{center}
\end{figure}

In the samples of class four the magnetization saturates at high $T$,
and the minimum value decreases with size (Fig.~\ref{magnet}). At
higher temperatures, 
between $T=3$ and 10, it increases again. This behavior we attribute
to the finite size of the samples.  The value of the maximum of the
susceptibility grows almost linearly with the generation $n$ and moves
to lower $T$ (Fig.~\ref{suscept}). It is not yet clear if the relation
$\chi_{max} (n) \propto (n-n_0)$ is valid for $n > 7$. If yes, this
would be a slow 
divergence (more precisely: $\chi_{max}$($N$) is about proportional to
$\sqrt[\displaystyle \tau^{3}]{N}$
where $N$ is the number of lattice
cells and $\tau$=$\frac{\displaystyle\sqrt{5}+1}{\displaystyle 2}$).  
With our current simulation programs it is not possible to
deal with $n\geq8$.

\begin{figure}
\caption{Binder order parameter. Sizes are indicated by the number of 
lattice points and the generation parameters $n$.\label{bop}}
\begin{center}
\includegraphics[width=13cm,angle=0]{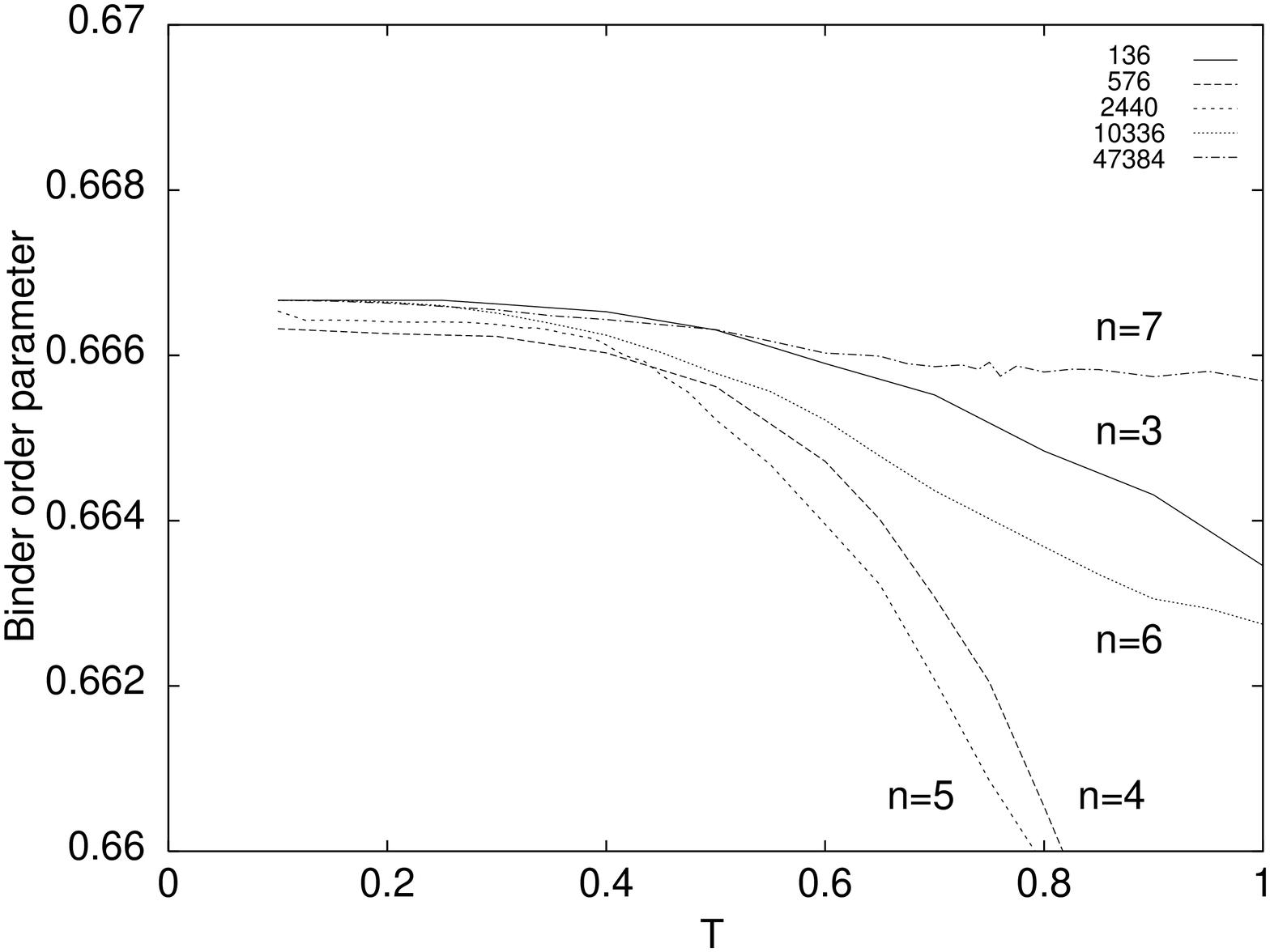}
\end{center}
\end{figure}

The Binder order parameter does not lead to a decision either,
since there is no clear intersection point visible
(Fig.~\ref{bop}). The trend of $B^{(N)}(T)$ changes at $n=5$, similar
to the magnetization behaviour. The most probable intersection point
is between $T=0.3$ and 0.4 if it exists.

But the tendency of $\chi_{max}$ with $n$ and the behaviour of
$B^{(N)}(T)$ indicate that a phase transition is much more likely to
occur at $T=0$.

\subsubsection{Self-diffusion coefficient}\label{sdc} 

\begin{figure}
\caption{Self-diffusion coefficient. Sizes are indicated by the number
of lattice points. \label{diffcoedd}}
\begin{center}
\includegraphics[width=9cm,angle=270]{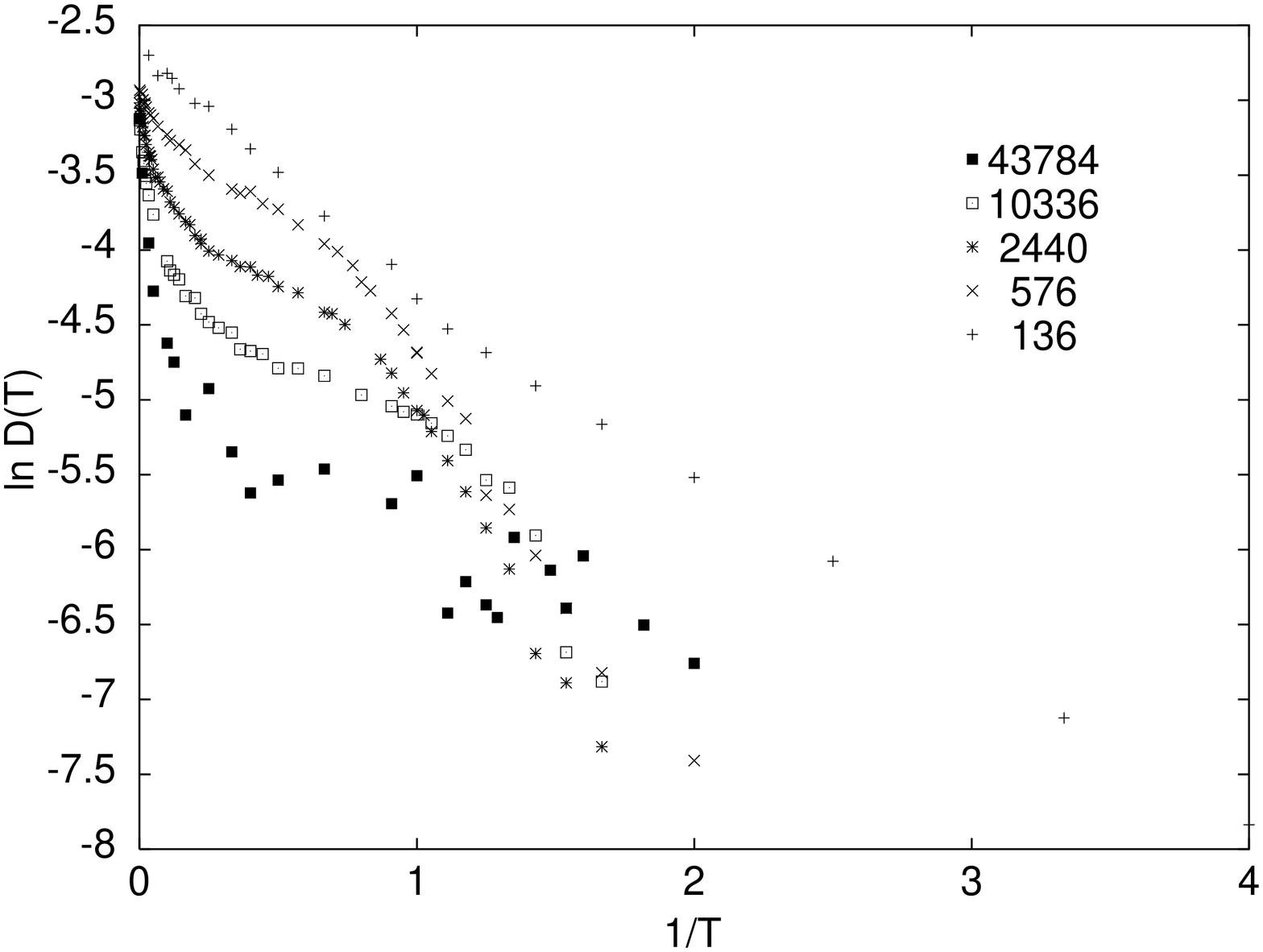}
\end{center}
\end{figure}

The mean square displacement $\left< ({\bf r}(t)-{\bf r}(0))^2
\right>$ grows linearly with time $t$, indicating a normal diffusion
behaviour and allowing the calculation of the diffusion coefficient
$D$. At temperatures lower than $T \approx 1$ the extrapolation is
difficult due to large fluctuations. $D$ becomes unmeasurably small
below $T=0.5$.  The diffusion coefficient forms a plateau at $T
\gtrapprox 1$ for $n=4,5,6$ in the \textsc{Arrhenius} plot
(Fig.~\ref{diffcoedd}). There may be several reasons for this behaviour:

First, there are energy barriers between different tiling configurations due
  to the harmonic energy measure, 
which at low temperatures lower the
mobility of lattice points for flips of higher energy.  In the range
of the plateau the probability for a flip only occasionally
suffices to overcome the barriers which play no role at high
temperatures.

Second, a phase transition may occur which changes the slope in
the \textsc{Arrhenius} plot.
G\"ahler has also observed a change of the slope in case of
the energy measure with the alternation
condition which turns out to be a phase transition since other
response functions like the susceptibility and the specific heat
definitely yielded a divergence at the same temperature. 

Last, there is an explanation which comes from the fact that the
number of flippable lattice points (number of simpletons) changes with
temperature.  We can distinguish four ranges:
\begin{enumerate}

\item The number of simpletons is about 23\% in the range $0\leq
  T\lessapprox 1$. 
  
\item In the range $1\lessapprox T \lessapprox 10$ we find a nearly
  logarithmic decrease of the number of simpletons. The plateau of $D(T)$
  is clearly seen here.

\item In the range $10\lessapprox T \lessapprox 100$ the approach of
  the frequency of the simpleton to a constant value leads to an
  increase of the negative slope of $D(T)$.

\item Above $T \gtrapprox 100$ the number of simpletons is constant at
  $\approx 17.5\%$. 
\end{enumerate}

In the approximants the behaviour of $D$($T$) is obscured to some
degree by zero energy modes 
caused by periodic boundaries.  These
modes become less and less important at larger sizes, but suppress the
plateau for small sample sizes.

Within the framework of the random tiling model we have studied only
flip diffusion. No other diffusion mechanism, in particular vacancy
diffusion, can be introduced in this way. The latter mechanism is
expected to be the dominant diffusion process at least above
600$^{\circ}$C, but there are indications that the new mechanism plays a
r{\^o}le below that temperature~\cite{bluher97}.

\subsection{Structure functions}

\subsubsection{Radial density function, radial structure factor, pair
interaction energy}\label{resrdf} 

\begin{figure}
\caption{Radial density function (RDF) of the rhombohedra lattice points. The
  peaks represent the values for the ideal tiling, the crosses denote the
  difference for the random tiling at $T=\infty$.\label{rdf}}
\begin{center}
\includegraphics[width=9.5cm,angle=270]{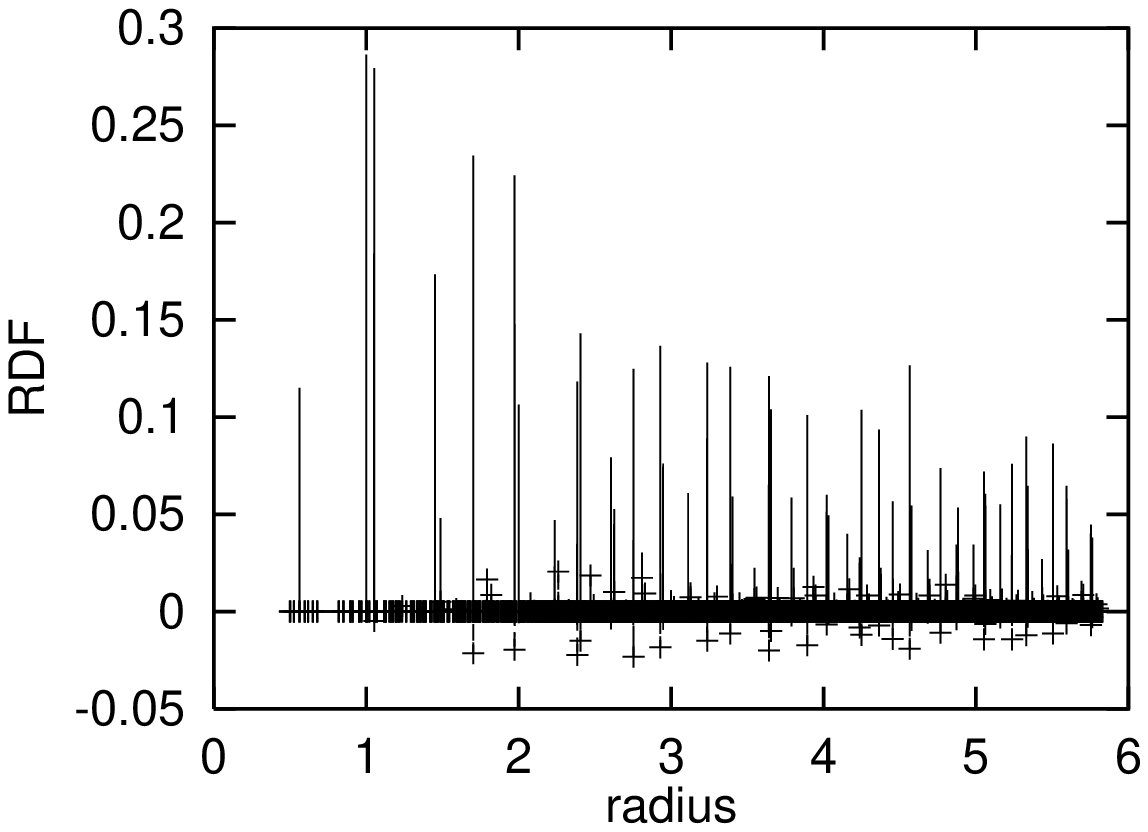}
\end{center}
\end{figure}

Structure functions have been calculated for samples of type four and
size $n=6$, containing 10336 lattice points, for the primitive 
monoatomic and the binary decoration~\cite{henley86}.
There are only small changes in the radial density function $g(r)$.
The largest differences of the order of 12\% occur for bonds between
two large atoms, but 
they are the least significant ones for stability due to their small
number. The energetically most important nearest-neighbour bonds
change even less in all cases, since a large portion of them lives
{\em within} the rhombohedra and thus is only transferred to another
position by a flip, but the frequency itself does not change.
The differences we observe in the addition tiling are somewhat larger,
but still very small. Figure \ref{rdf} shows the results for a
monatomic sample of size $n=5$.

The trend of the radial structure factor $I(k)$ which corresponds to the
\textsc{Fourier}-transformation of $g(r)$ is similar to that of the radial
density function. Due the smallness of the sample there are rather
large finite size effects which do not permit a quantitative comparison.

We further observe a very small dependence of the pair energy function
(see Eq.~\ref{pairpot}) on
size and $T$ due to the relation between the pair energy and the pair
distribution function. It is interesting to note that the potential
energy of the random tiling ensemble is at $T=\infty$ about 0.4\% lower than
for the ideal quasicrystal for a large class of
\textsc{Lennard-Jones}-like potentials, whereas the fluctuations
within the random tiling ensemble are of the order of 0.1\%.
Therefore simple pair potentials obviously favour energetical
equidistribution, i.~e.\ the random tiling model. 

\subsubsection{Variance, vertex statistics} 

The variance behaves roughly linear with respect to $u$ which
is obvious if one compares their definition in Eqs.~\ref{vareq} and
\ref{eneq}. The variance therefore shows a similar behaviour as the
internal energy.

The change in the statistics of the vertices with temperature is
much more important, as we have seen in the discussion of the diffusion
behaviour and in the r{\^o}le of the simpleton frequency in Sec~\ref{sdc}. 
The changes of the vertex frequencies (Notation as in~\cite{henley86b})
can be summed up as follows: 
\begin{enumerate}

\item The simpleton vertex 1 (452)\footnote{The first digit is the
    number of vertices connected by an edge, the second the number of
    vertices connected by a short face diagonal and the last the
    number of atoms connected by a short diagonal of the oblate
    rhombohedron} decreases from 23.61\% to 17.5\%.
\item Vertex 3 (670) decreases from 23.61 \% to less than 10\%.
\item Vertex 2 (561) increases from 23.61 \% to 29 \% at $T=8$ and then
  decreases to 26\%.
\item The sum of frequencies of vertex 1 and vertex 2 is roughly constant.
\item The forbidden vertices (all together) 
      increase from 0\% to more then 40\%.
\item The vertex with the highest symmetry where 20 prolate
  rhombohedra meet (vertex 24, in~\cite{henley86b} also called 
  twelve-fold sites)
  decreases in frequency from 1.2\% close to extinction.
\end{enumerate}

The addition tilings show a strong deviation from the ideal
distribution.  For the sizes $n=4,5,6$ we obtain for the simpleton
frequencies: 16.7 \%, 15.0 \%, 12.2 \%, and for the frequencies of the
forbidden vertices: 36.3 \%, 43.2~\% and 55.3\%, respectively.

We note that the statistical error of the vertex statistic is --- in the case
of equilibrium random tilings --- largely
independent of the sample size and already for $n$$\geq$4 very small. 

\subsubsection{Plane and axial structures}  

Plane and axial structures have been calculated for a sample of type
four with the size $n=6$, containing 10336 lattice points.  The area
function $A_s(\xi^{\perp})$ has been calculated from the dual lattice
points in the orthogonal space, and the histogram of the densities
$H(\rho_F)$ has been derived by numerical differentiation as described
in Sec.~\ref{plane}. The analysis has been carried out at 
temperatures $T=0.3$, 1, 3, $\dots$ 1000 for planes
perpendicular to two-, three- and fivefold planes.

There is a continuous transition from an area function
$A_s(\xi^{\perp})$ which is dependent on the symmetry directions to an
almost isotropic gaussian shape at 
high temperature, indicating that the acceptance domain
is smeared out isotropically.
At $T=\infty$ the histogram of densities $H(\rho_F)$ also becomes almost
independent of the symmetry direction. $H(\rho_F)$ is strongly peaked at
very low and very high plane densities which indicates that very
weakly and very densely occupied planes dominate.

In the five-fold direction we observe the smallest changes. Strong
differences between the  analytical and numerical functions,
especially at poles and singularities, for three- and
two-fold directions prohibit quantitative predictions.

The addition tilings show a broader distribution in
$A_s$($\xi^{\perp}$) and an anisotropy. Additional minor maxima occur
in $H(\rho_F)$, but they alternate from one realization to the other.
The distribution of dual lattice points
of one tiling is anisotropic by construction. To
obtain $A_s(\xi^{\perp})$ we have averaged over all
symmetry-equivalent directions.

We could not find any useful results for axial structures. The size of
the samples appears to be too small to give a satisfying resolution
for histograms. Strong differences between analytical and numerical
$H(\rho_l)$ functions for $T$=0 also indicate that the results are not
reliable.

\subsubsection{Diffraction patterns} 

Diffraction patterns (Bragg scattering without observation of diffuse
line shapes which are characteristic for random tilings) have been
calculated for a sample of type four with the size $n=5$, containing
2440 lattice points. The analysis has been carried out at 
temperatures $T=0$ 0.3, 1, 3, 10, and at $T=10000$ for planes
perpendicular to two-, 
three- and fivefold planes.  The intensities are only weakly dependent
on temperature as the acceptance domain is only weakly smeared out
even at high temperatures. This result is consistent with Tangs
observation of limited phason fluctuations in three
dimensions~\cite{tang90}. 

In the case of the addition tilings stronger differences exist: Some
of the reflexions are weaker, and new ones arise. This indicates that
there is already a large phasonic disorder in the addition tilings
which generates an extended cloud of dual lattice points.

\section{Conclusions}
\label{conclude}

We have investigated a geometric and thermodynamic
model of disordered quasilattices,
which could explain the structure of icosahedral metal alloys.

In this random tiling model there are no matching rules, in contrast
to the ordered quasiperiodic structures. The only requirement is that the
physical space must be filled with rigid tiles without gaps or
overlaps.

The results are summarized as follows:

\begin{enumerate}
\item The internal energy $u$ increases and saturates at high
  temperatures. $u_{\infty}$ converges towards a limit for
  $N$$\to$$\infty$, which can be estimated from values of the phason
  fluctuation variance for the pure entropic random tiling ensemble.
  The entropy $s_{\infty}$ obtained from the specific heat $c_V$
  by thermodynamic integration is about 0.24$\pm$0.01, in agreement
  with results from many other sources~\cite{strandburg91,jaric95,gaehler95}.
  The specific heat $c_V$ shows a \textsc{Schottky} anomaly. It is
  not clear whether it diverges. The increase of the entropy with the
  system size comes from the broadening of the maximum.

\item The sheet magnetization decreases to a minimum with the temperature
  and slightly grows to a saturation value. In the thermodynamic
  limit it should vanish without an intermediate minimum.
  The susceptibility diverges slowly, the maximum shifts to
  smaller temperatures.
  The Binder order parameter does not exhibit a unique
  intersection point.  We assume that the transition temperature is
  finite, hence the random tiling transition could be analogous to
  the transition in a two-dimensional Ising ferromagnet.
  
\item The self-diffusion coefficient displays a plateau in the central
  temperature range indicating energy barriers for certain flips due
  to the harmonic energy measure. On the
  other hand there exist correlations between the temperature
  dependence of the self-diffusion coefficient and the frequencies of
  simpletons per lattice point. The Arrhenius plot is deviating
  strongly from that assumed by Kalugin and Katz~\cite{kalugin}: Where these
  authors are plotting a steep increase, we are observing the
  plateau. There are zero energy modes due to
  periodic boundary conditions, which lead to a suppression of the
  plateau for small approximants.
  
\item Henleys postulate of the finiteness of the phason fluctuations
  is demonstrated by the Monte-Carlo-simpleton-flip tilings. The
  addition tilings show larger variances and deviations of the vertex
  frequencies from the equilibrium values. There are also changes in
  the diffraction patterns caused by phason fluctuations.
  
\item The frequencies of the densities of planes in the equilibrium
  tilings are temperature dependent and very weakly and very densely
  decorated planes dominate. Addition tilings show further minor
  maxima, dependent on the realization.
  
\item Radial structure functions depend only weakly on the
  configuration. This is due to the rigidness of the cells. Pair
  interactions realize an equidistribution of all configurations --- a
  possible realization of the pure entropic random tiling model.

\end{enumerate}

The behaviour of $c_V$ and $\chi$ obviously depends strongly on the
type of energy measure used. The alternation condition seems to work
much better~\cite{gaehler95} --- maybe as consequence of its closer
similarity with Ising interaction models in comparison with the
harmonic energy measure. However, for the simple energy
model~\cite{joseph95} it was also not possible to decide if a phase
transition occurs, since no divergence of the specific heat could be
observed.

\section*{Acknowledgements}

The authors are very indebted to Franz G\"ahler for helpful
discussions.

\begin{table}
\begin{center}
\caption{Configurational entropy for triacontahedral clusters with
fixed surface. The first column codes the type of polyhedron by the 
length of edges in units of the six icosahedral basis vectors. The
second column gives the number of configurations, the third the number of
rhombohedra. The last column contains the configurational entropy per
lattice point.
\label{tabent}}
\begin{tabular}{c|r|c|c}
edge lengths & \# configs. & \# rhombo. & s \\
\hline
111111 & 160 &20 &0.2538\\
211111 & 1280 &30 &0.2385\\
112211 & 22381 &44 &0.2276\\
121121 & 22981 &44 &0.2282\\
221121 & 1268131 &63 &0.2231\\
\end{tabular}
\end{center}
\end {table}

\begin{table}
\caption{Some thermodynamic parameters for the harmonic interaction
  model with periodic boun\-da\-ry conditions. The first column provides the
  generation for cubic approximants, the se\-cond column the number of
  lattice points,
  the following columns list the entropy $s_{\infty}$ and
  the variance $\Omega$($T$=$\infty$). The rows ``890'' and ``1440'' denote
  pentagonal approximants.
\label{tab1}}

\begin{center}
\begin{tabular}{c|c|c|c}
$n$ & size & $s_{\infty}$ & $\Omega(T=\infty)$\\
\hline
3 & 136 & 0.1816878& 1.475 $\pm$ 0.015\\
4 & 576 & 0.2032964& 1.56 $\pm$ 0.01\\ 
& 890 & 0.2086351&1.60$\pm$ 0.02\\
& 1440 & 0.2076728&1.615$\pm$ 0.02\\
5 & 2440 & 0.2114427 & 1.635 $\pm$ 0.01\\
6 & 10336 & 0.2349887 & 1.675 $\pm$ 0.01\\
7 & 43784 & 0.2367582 & 1.70 $\pm$ 0.01\\
\hline
 & limit & 0.24$\pm$0.01 & 1.73 $\pm$ 0.01
\end{tabular}
\end{center}
\end{table}

\begin{figure}
\begin{center}
\caption{Simpleton flip\label{flip}}
\vspace*{1cm}
\includegraphics[width=14cm,angle=0]{simpleton.eps.gz}
\end{center}
\end{figure}

\begin{figure}
\begin{center}
\caption{Worm of rhombohedra\label{worm}}
\vspace*{2cm}
\hspace*{-2cm}\parbox{16cm}{\includegraphics[width=12cm,angle=0]{3D-Baender.eps.gz}}
\end{center}
\end{figure}

\newpage

\newpage

\newpage

\newpage

\end{document}